# Bandgaps and directional propagation of elastic waves in two-dimensional square zigzag lattice structures

Yan-Feng Wang[1], Yue-Sheng Wang[1,*] and Chuanzeng Zhang[2]

[1]Institute of Engineering Mechanics, Beijing Jiaotong University, Beijing 100044, China

[2]Department of Civil Engineering, University of Siegen, Siegen 57068, Germany

In this paper we propose some kinds of two-dimensional square zigzag lattice structures and study their bandgaps and directional propagation of elastic waves. The band structures and the transmission spectra of the systems are calculated by using the finite element method. The effects of the geometry parameters of the 2d-zigzag lattices on the bandgaps are investigated and discussed. The mechanism of the bandgap generation is analyzed by studying the vibration modes at the bandgap edges. Multiple wide complete bandgaps are found in a wide porosity range owing to the separation of the degeneracy by introducing bending arms. The bandgaps are sensitive to the geometry parameters of the systems. The deformed displacement fields of the transient response of finite structures subjected to harmonic loads are presented to show the directional wave propagation. The research work in this paper is relevant to the practical design of cellular structures with enhanced vibro-acoustics performance.

## 1. Introduction

Extensive efforts have been exerted on the study of wave propagation in phononic crystals (PCs) [1], a kind of artificial structures with periodic varying elastic properties and mass densities. Particular interests are focused on the unique characteristic of PCs, i.e., the bandgap, within which the propagation of elastic waves is forbidden. Thus, the related study is of interest for innovative applications in vibrationless environments for high-precision mechanical systems, filtering and negative refraction [2].

In recent years, many works have been devoted to the study of PCs with periodic holes [3-12], the bandgaps of which are mainly determined by the geometry parameters. The cross section of the holes may be either convex [3-6, 9-11] or non-convex (e.g. cross-like [7]). A system composed of large lumps connected by small connectors is easy to generate complete bandgaps due to the local resonances of the lumps. Such systems can be obtained by etching cross-like holes, circular holes, or rotated square holes in the matrix [3, 5, 7-9, 12]. The porosity (i.e. the filling fraction of the vacuum part), $f$, of such systems generally has some intermediate values, and the porosity range for the existence of the bandgaps is small. Other systems without large lumps can hardly exhibit complete bandgaps [4], especially in the low frequency range [6, 9, 11]. For lattice structures with extremely high porosity ($f \rightarrow 1$) [13-19], which are widely used as stand-along structural components, narrow complete bandgaps exist for particular lattice forms and/or when the slenderness ratio of the beam is extremely big [15]. So it is a big challenge to find wide complete bandgaps in the low frequency range for PCs with a high porosity. In spite of bandgaps, directional propagation of waves in the lattice structures is also a topic which has received considerable attention [13, 14, 20, 21].

* Corresponding author, E-mail: yswang@bjtu.edu.cn, Fax: +86-10-51682094

In this paper, we study the wave propagation behaviors of a kind of lattice structures with zigzag arms [22, 23]. Such structures can generate complete bandgaps in a wide porosity range even when the porosity is very small. Compared to the system with straight arms, they can generate complete bandgaps at lower frequencies, and also present good directional properties. Numerical simulations are implemented by using the finite element method. Both band structures and transmission spectra of the systems are calculated. The effects of the geometry parameters on the bandgaps are investigated and discussed. The vibration modes at the bandgap edges are calculated to analyze the mechanism of the bandgap generation. The phase constant surfaces of some typical bands, as well as the transient propagation of elastic waves in finite structures, are calculated to show the directional properties of the considered systems.

## 2. Problem statement and computational model

We consider two kinds of two-dimensional square zigzag lattice structures, see Fig.1. They have zigzag arms in one [Fig. 1(a)] or two [Fig. 1(b)] directions. We simply name them as 1d- and 2d-zigzag lattices, respectively. Suppose that the lattice constant is $a$, the bending angle of the zigzag arm is $\theta$, and the horizontal (or vertical) distance between two inflection points is $l$, see Fig. 1. Then the other geometry parameters shown in the figure can be determined from the above parameters. For instance, the widths of the arms are $d_0 = a(1-\sqrt{f})$, $d_1 = d_0 \cos\theta_1$ and $d_2 = d_0 \cos(\theta - \theta_1)$; and the rotational angle of the zigzag arm is

$$\theta_1 = \arctan \frac{\sqrt{a^2 + 4l(a-l)\tan^2\theta} - a}{2l\tan\theta}.$$

Typically, when $l/a$=0.5, we have $\theta_1 = \theta/2$, and thus

$$d_1 = d_2 = a(1-\sqrt{f})\cos(\theta/2), \tag{1}$$

which indicates that the zigzag arms have an identical width. When $\theta = 0^\circ$, we have the square lattice structure with straight arms.

An alternative profile of the 2d-zigzag lattice may be obtained by perforating a solid with rotational cross holes [7], see Fig. 3(a). The cross-section of the unit cell of this alternative profile is shown in Fig. 3(b), where $a$ is the lattice constant, $\varphi$ is the rotational angle of the hole, while $b$ and $c$ denote the other geometry sizes of the cross hole, respectively. The rotational angle of the cross hole is limited to the range of $\arcsin(c/a) < \varphi < \arcsin[b/(2a)]$ ($c<a<b<2a$). For simplicity, we name the two parts between the inflection points as the main and the secondary arms, see Fig. 3. Their arm widths are $(2a\cos\varphi - b - c)/2$ and $(a\sin\varphi - c)$, receptively. This structure has the same rotational symmetry as the one in Fig. 1(b).

In this paper, the bandgap properties and directional wave propagation in the systems shown in Figs. 1 to 3 will be studied with emphasis on the effects of the bending arms. The finite element method based on COMSOL 3.5a [7, 19] is used to calculate the band structures of the considered systems. Due to the periodicity, only one unit cell is used in the calculation. In addition to the antisymmetric zigzag lattices shown in Fig.1, we also consider the symmetric systems because the difference between the symmetric and antisymmetric lattices may lead to different wave propagation properties. The finite element models of the symmetric unit cells and the associated irreducible Brillouin zones are depicted in Fig.2. The default triangular mesh with Lagrange quadratic elements provided by COMSOL is used with a double refinement. Bloch conditions [7] are applied on the opposite boundaries of the unit cell, and traction-free boundary conditions are applied on the other boundaries. The whole dispersion relations can be obtained by sweeping the wave vector $\mathbf{k}$ along the edges of the irreducible Brillouin zone. A detailed description of the procedure can be found in Ref. [7]. It is worthy to be noticed that the irreducible Brillouin zones are different for different unit cells, see Figs. 1 and 2.

The wave transmission properties of the systems are evaluated by using the transmission coefficients. The configuration for the calculation is shown in Fig. 4. Perfectly matched layers (PMLs) [24] are used as extended domains at the artificial model domain boundaries to eliminate the unphysical wave reflections. PMLs have exactly the same elastic property and mass density as the matrix, but have an artificial gradient damping coefficient along the wave propagation direction. Consequently, they will not disturb the propagation of the elastic waves, emulating an infinite media.

In general, the mixed plane wave modes have co-existing horizontal ($u$) and vertical ($v$) displacements. So, two different incident waves (i.e., the $x$- or $y$-polarized wave) are generated by applying a displacement field of the unit amplitude on the left side of the system. Periodic boundary condition is applied on the $y$-direction, which emulates an infinitely long line source. The amplitude of the transmitted waves is defined as the average amplitude of all the nodes on the observation line in the right part of the system. Then, the transmission coefficient with the unit of dB is defined as

$$\log\left(\frac{\int_s (|U_t|/L)\mathrm{d}s}{|U_i|}\right), \tag{2}$$

where $|U_t|$ and $|U_i|$ are the amplitudes of the transmitted and incident waves, respectively; and $s$ is the observation line with the length of $L$, where the transmitted wave is received.

## 3. Mechanism of the bandgap generation

This section will present some detailed numerical results for various zigzag lattices and analyze the mechanism of the bandgap generation. The elastic parameters of the aluminum matrix (shadowed parts in Figs. 1 and 2) are $\rho = 2700\mathrm{kgm}^{-3}$, $E = 20\mathrm{GPa}$ and $\nu = 0.25$ [7]. For convenience, the reduced frequency $\Omega = \omega a/(2\pi c_t)$ is introduced, with the lattice constant $a = 0.02\mathrm{m}$ and the transverse wave velocity of the matrix $c_t = 1721\mathrm{m/s}$. The white parts in Figs. 1 and 2 represent vacuums.

The dispersion curves for the 1d- and 2d-zigzag lattices with $\theta = 40^\circ$ are shown in Fig. 5. For comparison, the results for the system with straight arms ($\theta = 0^\circ$) are presented in Fig. 5(a). Transmission spectra for the above systems are also presented to validate the calculation of the band structures. If both the *x*- and *y*-polarized waves are attenuated clearly in the transmission spectra in a particular frequency range, then we get a directional bandgap in the corresponding direction. The directional bandgaps predicted by the band structures and the spectra are highlighted by the shadowed regions in the figures. The results predicted by the transmission spectra and the band structures are in good agreement.

It is noted that no complete bandgap is observed in Fig. 5(a) for the system with straight arms, as demonstrated in Ref. [7]. While for the antisymmetric 1d-zigzag lattice in Fig. 5(b), multiple directional bandgaps rather than complete bandgaps appear dominantly in the bending arm direction, but barely in the straight arm direction. In the symmetric 1d-zigzag lattice the directional bandgap between the 6th and 7th bands gets wider, and one narrow complete bandgap appears in the frequency range of $0.67 < \Omega < 0.71$, see Fig. 5(c). However, three complete bandgaps are observed in the antisymmetric 2d-zigzag lattice, as shown in Fig. 5(d). The lowest one exists between the 4th and 5th bands in the frequency range of $0.35 < \Omega < 0.45$. For the symmetric 2d-zigzag lattice in Fig. 5(e), one complete bandgap between the 6th and 7th bands is observed in the frequency range of $0.51 < \Omega < 0.78$, which is wider but higher than that of the antisymmetric structure. And the bandgap width is about 7 times wider than that of the symmetric 1d-zigzag lattice shown in Fig. 5(c).

The above results imply that a system with bending arms is favorable to open multiple wide directional or complete bandgaps at relative low frequencies. Bending arms in two directions can open wide complete bandgaps easilier than those in one direction. For the system with straight arms, no complete bandgap except a few directional bandgaps exists. This is because there are many crossover regions [25] which occur either on the high symmetry points [e.g., B, G and H in Fig. 5(a)] where folding of bands is expected due to the periodicity, or inside the first Brillouin zone [e.g., D, E, F, I and J in Fig. 5(a)]. Different from the straight arm lattice, the symmetric (or antisymmetric) zigzag lattice possess reflection (or central) symmetry. Different types of symmetry are the reason for the degeneracy lifting, and resulting in new directional or even complete bandgaps. For instance, the degeneracy at the crossover region I is separated for either symmetric or antisymmetric 1d-zigzag lattice in both the bending arm direction and the straight arm direction, which directly gives rise to a new directional bandgap between 9th and 10th bands. In the antisymmetric 1d-zigzag lattice [Fig. 5(b)], the new directional bandgap appearing in the $\Gamma M$ (or $\Gamma M'$) direction between the 5th and 6th bands is owing to the separation of the degeneracy at the crossover regions C and D. Furthermore, the degeneracy at the crossover regions D, E and F is separated in the bending arm direction, but not in the straight arm direction, and therefore resulting in the bandgaps in the $\Gamma X$ direction, see Fig. 5(b). However, for the symmetric 1d-zigzag lattice in Fig. 5(c), the degeneracy at the crossover region D (but not at the crossover region E) is separated only in the bending arm direction, leading to the bandgap in the $\Gamma X$ direction.

Actually, the crossover regions D, E and I in Fig. 5(a) are real cross points, i.e., the dispersion curves are the true intersection [26]. As defined in Ref. [27], the amount of the polarization along the *x*-axis can be represented by one positive number, and given by

$$\frac{\int_S |u|^2 \,\mathrm{d}S}{\int_S (|u|^2 + |v|^2)\,\mathrm{d}S}, \tag{3}$$

with the integral taken over the whole unit cell *S*. The amount of the polarization along the *x*-axis is 0.97 at point $B_1$, and 0.06 at point $B_2$. This means that these two modes are dominantly *x*- and *y*-polarized, respectively. Similar behavior is observed at the crossover regions C and D. While the amount of the polarization along the *x*-axis is 0.5 at point $E_1$, and 0.99 at point $E_2$. So the coupling effect between the two modes at the crossover region E is stronger than that at the crossover regions B, C, and D. For both symmetric and antisymmetric 1d-zigzag lattices, the weak degeneracy at the crossover regions B, C and D is separated in the bending direction. While only the antisymmetric 1d-zigzag lattice separates the relatively strong degeneracy at the crossover region E. So the weak degeneracy is easy to be separated when the arms are bended; and the decrease of the symmetry of the system can result in the separation of a relatively strong degeneracy. The amount of the polarization along the *x*-axis is 0.97 at point $I_1$, and 0.46 at point $I_2$. The coupling effect between the modes at the crossover region I is nearly the same as that at the crossover region E. While the degeneracy at the crossover region I is easier to be separated because it exists at a higher frequency compared to that at the crossover region E.

In the 2d-zigzag lattices, the bending effects are extended into two directions, and the effect of the degeneracy separation generally gets stronger. The directional bandgaps (e.g., the ones generated owing to the separation of the degeneracy at the crossover region I) get wider compared to those for the 1d-zigzag lattices. In particular, multiple wide complete bandgaps appear in the band structures. For instance, in the antisymmetric 2d-zigzag lattice, the lowest bandgap occurs owing to the separation of the degeneracy at the crossover region E, see Fig. 5(d). It should be pointed out that, by introducing the antisymmetric 1d-zigzag lattice, the natural mode at point $X_4^0$ changes from Fig. 6(a) to Fig. 6(b) [point $X_4^1$], the eigenfrequency of which is smaller than that at point $X_4^0$, and results in the decrease of the 4th band in the bending arm direction. Thus, the degeneracy at the crossover region E is separated, even if no new directional bandgap appears. For the antisymmetric 2d-zigzag lattice, the corresponding mode changes to be Fig. 6(c) [point $X_4^2$], the eigenfrequency of which gets even small. The 4th band decreases in all the bending arm directions, and gives rise to the lowest complete bandgap. In the symmetric 2d-zigzag lattice [Fig. 5(e)], the separation of the degeneracy at the crossover region B gets more distinct compared with the symmetric 1d-zigzag lattice in Fig. 5(c). This pronounced effect is because of the larger decrease of the eigenfrequencies from Fig. 6(d) [point $M_6^0$] to Fig. 6(f) [point $M_6^2$] in comparison with that from Fig. 6(d) to Fig. 6(e) [point $M_6^1$]. Then the corresponding band decreases and the associated complete bandgap becomes wide. These results also imply that different symmetry can result in different vibration modes, and consequently different bandgaps. However, in some cases, the effect of the degeneracy separation gets weaker. For example, a directional bandgap is generated owing to the degeneracy separation at the crossover region G for the 1d-zigzag lattice [Figs. 5(b) and 5(c)]. However, this degeneracy is just separated for the symmetric 2d-zigzag lattice [Fig. 5(e)], but not for the antisymmetric one [Fig. 5(d)].

The above results indicate that complete bandgaps can be induced due to the separation of the degeneracy when the arms are bended. Different types of the symmetry can give rise to different directional or complete bandgaps. The generation of the lowest bandgap is owing to the increase of the frequency of the rotational mode of the unit cell. The appearance of the modes with incomplete standing still points or big vibrationless parts narrows or eventually closes the bandgap.

The dispersion curves for the perforated 2d-zigzag lattice with cross holes are shown in Fig. 7(a), where the rotational angle of the cross holes is $\varphi = 30^{\circ}$. Similar to the 2d-antisymmetric zigzag lattice, this system also presents multiple complete bandgaps. The lower edge mode of the lowest bandgap is shown in Fig. 7(b), the eigenfrequency of which is smaller than those in Figs. 6(a)-6(c). So this system can generate a complete bandgap at a lower frequency.

It is also noted that the mechanism of the bandgap generation for the proposed system is different from that for a system with circular holes, where the locally translational resonance of the large lumps dominates the appearance of the complete bandgaps [7]. For the present system, the bandgaps are generally induced by the rotational vibration which is mostly more complicated than the translational resonance. Because of more freedom in selecting geometrical parameters, the bandgaps of the present system are easily engineered. Furthermore, it will be proved later that the proposed system can exhibit complete bandgaps in a wider porosity range at lower frequencies compared to the system with circular holes.

## 4. Effects of the geometry parameters on bandgaps

In this part, we will mainly discuss the effects of the geometry parameters on complete bandgaps for the antisymmetric 2d-zigzag lattice (Fig. 1b) and the perforated 2d-zigzag lattice with cross holes (Fig. 3).

### 4.1 The antisymmetric 2d-zigzag lattice

Band structures of the antisymmetric 2d-zigzag lattices with different bending angles are illustrated in Fig. 8(a). It is shown that no complete bandgap appears in the case of straight arms, i.e., when $\theta = 0^{\circ}$. When the arms are slightly bended ($\theta = 10^{\circ}$), the degeneracy at the crossover region C is separated and one complete bandgap comes up in a relatively high frequency range. The degeneracy at the crossover regions A and B is also separated. When $\theta = 30^{\circ}$, the lowest complete bandgap appears due to the increase of the 6th band. The upper edge mode changes from point A in Fig. 8(b) to point B in Fig. 8(c) when $\theta = 40^{\circ}$. These two modes have a rotational vibration. However, there are four complete standing points at the ends of the arms in Fig. 8(b) but four incomplete standing points in the middle parts of the arms in Fig. 8(c). Thus these two modes will play different roles. The former one will result in the generation of the bandgap, while the latter one can narrow the bandgap. Keeping on increasing the bending angle, some bandgaps in a high frequency will appear again.

To further show the dependence of the geometry parameters on the bandgaps of the antisymmetric 2d-zigzag lattice, variations of the bandgap edges with the geometry parameters are demonstrated in Fig. 9. It is shown that multiple complete bandgaps appear in the systems. The first (lowest) one appears between the 4th and 5th bands, the second one between 6th and 7th bands, the third one between 8th and 9th bands, and the fourth one between 9th and 10th.

The bandgap edges varying with different bending angle $\theta$ are plotted in Fig. 9(a) with the fixed porosity $f$=0.8 and distance $l$=0.5$a$. It is shown that no complete bandgap appears when $\theta = 0^{\circ}$. The first (lowest) bandgap comes up between the 4th and 5th bands when $\theta = 26^{\circ}$. With the increase of the bending angle, the upper edge first increases due to the increase of the natural frequency of the mode at point A. When $\theta > 40^{\circ}$, the upper edge mode changes to be similar to that at point I, the natural frequency of which decreases with the increase of the bending angle. While the lower edge, or the natural frequency of the vibration mode at point B, decreases monotonously. The bandgap width first becomes big and then small with the decrease of the bending angle. The second complete bandgap first exists when $24^{\circ} < \theta < 44^{\circ}$ with a very small width. When $\theta > 50^{\circ}$, it comes up again. The variations of its bandgap edges and the bandgap width are similar to those of the first complete bandgap. Similar to the case of the second complete bandgap, the third complete bandgap first exists at relatively high frequencies when $4^{\circ} < \theta < 34^{\circ}$ with a small bandgap width. When $\theta > 80^{\circ}$, it comes up again, with the bandgap edges decreasing with the bending angle increasing. The fourth complete bandgap appears when $\theta = 40^{\circ}$. Its bandgap edges vary in a similar trend to the first bandgap with the increase of the bending angles.

Fig. 9(b) illustrates the bandgap edges varying with the porosity $f$ for the fixed arm width $d_1 = d_2 = d = 0.09a$ and distance $l$=0.5$a$. Four complete bandgaps appear when the system has a relatively high porosity (or equivalently when the bending angle determined by Eq. (1) is relatively big for the fixed arm width). With the porosity decreasing, the widths of all the bandgaps first become wide and then get small. The lower edges of all these bandgaps decrease monotonously. While the upper edges first increase and then decreases except for the third bandgap. Indeed, the systems with circular holes can also generate complete bandgaps, but within a small porosity range ($0.5 < f < 0.79$) [28]. And only one geometry parameter, namely the diameter of the holes can be used to tune the bandgaps. Moreover its lowest bandgap edge, which can be obtained when its porosity reaches the limiting value $f$=0.78, is about $\Omega = 0.3$. However, the lower edge of the first bandgap is about $\Omega = 0.2$ for the proposed system when $f$=0.78, and it can be easily tuned to a smaller value. So the proposed system shows some merits in practical design.

Variations of the bandgap edges with different porosity $f$ for the fixed bending angle $\theta = 60^{\circ}$ and distance $l$=0.5$a$ are presented in Fig. 9(c). The four complete bandgaps exist in a wide porosity range from a small to a big value approaching to 1 (or equivalently when the arm width determined by Eq. (2) varies in a wide range for the fixed bending angle). The width of all the bandgaps first become big and then get small. The second and third bandgaps disappear when the porosity has an intermediate value. With the increase of the porosity, all these bandgap edges decrease monotonously, except the upper edge of the third one.

In Fig. 9(d), the bending angle $\theta$ and the porosity $f$ are fixed, then the horizontal (vertical) distance $l$ is no longer a constant, and thus the arm width will be various. No complete bandgap appears when the distance $l$ is small. The first three complete bandgaps come up when $l/a$=0.3, 0.27 and 0.22, respectively. With the increase of the bending distance, their lower edges first decrease and then increase. While their upper edges first increase and then decreases, except the upper edge of the second one which increases monotonously. All these bandgaps obtain their maximum width when $l/a$ takes certain optimal values.

**4.2 The perforated 2d-zigzag lattice with cross holes**

Figure 10(a) presents the band structures for different rotational angles of the cross holes. It is shown that no complete bandgap appears in the first ten bands when $\varphi = 23.68^{\circ}$. When $\varphi = 23.7^{\circ}$, one small complete bandgap appears between the 8th and 9th bands due to the increase of the $l_1$- and $l_2$-bands. The vibration mode at point B is shown in Fig. 10(b), where the shape of the cross holes may not be easily recognized because the secondary arm is very thin. Both the main and secondary arms vibrate, and the whole system has a rotational vibration, similar to those shown in Fig. 8. The natural frequency of the system is dominated by the vibration of the secondary arm. With the increase of the rotational angle, the width of the secondary arm increases, and thus the natural frequency increases. When $\varphi = 23.8^\circ$, one complete bandgap between the 6th and 7th bands comes up due to the increase of the degenerate $l_{4,5}$-bands. A complete bandgap between the 4th and 5th bands appears when $\varphi = 24.13^\circ$ as a consequence of the increase of the degenerate $l_{6,7}$-bands. When $\varphi = 25^\circ$, a complete bandgap appears between the 7th and 8th bands due to the increase of the $l_8$-band. The vibration mode of the $l_1$-band changes from Fig. 10(b) to Fig. 10(c) when $\varphi = 28^\circ$. With the increase of the rotational angle, the natural frequency of this mode decreases because the width of the main arm gets smaller. When $\varphi = 32^\circ$, the vibration mode of the $l_8$-band changes and this band decreases with the increase of the rotational angle.

Generally speaking, the generation of the complete bandgaps is attributed to the rotational vibration of the whole unit cell, as the antisymmetric 2d-zigzag lattice we discussed in the last section. When the vibration of the main arm plays a dominant role, the natural frequency of the mode decreases with the increase of the rotational angle due to the decrease of the main arm width, and thus results in the descent of the corresponding bandgap edges.

Variations of the bandgaps with the rotational angle or the size of the cross hole are shown in Fig. 11. The first (lowest) complete bandgap locates between the 4th and 5th bands, the second one between the 5th and 6th bands, the third one between the 6th and 7th bands, and the fourth one between the 7th and 8th bands.

Figure 11(a) plots the variation of the bandgaps with the rotational angles $\varphi$ for the fixed size of the cross hole ($b/a$ =1.2, $c/a$ =0.4). No complete bandgap exists when the rotational angle is of a small or big value. Complete bandgaps appear when the rotational angle is of some intermediate values. The third bandgap exists in a small range of $\varphi$ ($23.79 < \varphi < 24.9^\circ$), and is very narrow. The first and fourth bandgap first enlarges, then narrows and disappears finally. With the increase of the rotational angle, the upper edge of the first bandgap increases first due to the increase of the $l_{6,7}$-bands. Then it begins to decrease due to the change of the edge mode from Fig. 10(b) to Fig. 10(c). The lower edge nearly does not change when $\varphi < 30^\circ$, and then decreases with the increase of the rotational angle due to the decrease of the $l_3$-band. The upper edge of the fourth bandgap first increases and then decreases. The lower edge first increases with the rotational angle increasing till $\varphi = 26^{\circ}$; then it begins to decrease due to the change of the edge modes from point G to point J. When $\varphi = 29^{\circ}$, the lower edge mode changes once more, and the lower edge tends to increase with the rotational angle again. When $\varphi = 24.9^{\circ}$, the second bandgap appears, with its upper and lower edges first increasing and then decreasing with the increase of the rotational angle, owing to the increase/decrease of the $l_{6,7}$- and $l_8$-bands. And the width of this bandgap gets big first, latter small, and then big again. The existence of this wide bandgap when $\varphi$ is near $36^{\circ}$ is due to the resonance of the main arms which are quite thin in this case.

Variation of the bandgap edges with the geometry size $b/a$ for the fixed $c/a$=0.4 and rotational angle $\varphi = 30^{\circ}$ is presented in Fig. 11(b). All the bandgap edges decrease monotonously with the increase of $b/a$. The first, second and forth complete bandgaps vary in a similar trend with $b/a$. They fist enlarge, then narrow, and disappear finally. Similar to the second complete bandgap in Fig. 11(a) when $\varphi$ is big, the third complete bandgap exists when $b/a$ is small due to the resonance of the thin main arms. With the increase of $b/a$, this bandgap becomes narrow and disappears finally.

The bandgap edges varying with the geometry size $c/a$ for the fixed $b/a$=1.2 and rotational angle $\varphi = 30^{\circ}$ are illustrated in Fig. 11(c). The bandgap edges decrease monotonously with the increase of $c/a$ except for the upper edge of the first complete bandgap, which first increases slightly and then decreases, and the lower edge of the third bandgap (which increases slightly and monotonously). The variation of the bandgap width with $c/a$ is similar to that with $b/a$ in Fig. 11(b). The existence of the complete bandgap when $c/a$ is of a small or big value is also owing to the local resonance of a system with large lumps connected with narrow connectors [7].

In general, the rotational vibration mode plays an important role on the generation/close of the bandgaps. The bandgaps are sensitive to the geometry parameters of the zigzag lattice, which is quite useful in the practical design of the lattice structures.

## 5. Directional wave propagation

Next we will demonstrate that the proposed structures can exhibit inspiring directional wave propagation.

### 5.1 The 1d- and 2d-zigzag lattices

We first consider the symmetric 1d-zigzag lattice which has different configurations along the various directions. The 6th phase constant surface is chosen to estimate the direction of the elastic wave propagation. It is well known that the direction of the elastic wave propagation is identical to that of the group velocity. In a two-dimensional system, the group velocity is given by [29]

$$c_{gx} = \frac{\partial \omega}{\partial k_x}, \quad c_{gy} = \frac{\partial \omega}{\partial k_y}, \tag{4}$$

where $c_{gx}$ and $c_{gy}$ are the group velocity components along the $x$- and $y$-axis, respectively. Equation (4) implies that the direction of the group velocity can be evaluated by calculating the normal to the isofrequency curves in the phase constant surfaces. In our numerical computation, 120 rays from the contour center are supposed to be evenly circularly distributed. The cross points between the contour and the 120 rays are chosen. Then the direction of the elastic wave propagation can be described by evaluating the angle of the corresponding normal.

Isofrequency curves are shown in Fig. 12(a). For a wide frequency range, the curves are noncircular, which implies generally that the direction of the elastic wave propagation does not coincide with the wave vector. Similar results can be found in the grid-like structures [13, 14, 20]. When the vibration is excited in a given region in a grid structure, the displacement amplitude or the energy flux will get high in some directions and low in the others with respect to the average. It is also noted here that the isofrequency curve of a particular mode in a particular direction in Fig. 12(a) is open, which means that this mode cannot propagate in this direction. This phenomenon is in agreement with the directional bandgap along the $\Gamma X$ direction in Fig. 5(c).

The polar plot of the directional propagation of the elastic wave [20, 21], the angle of the normal to the isofrequency curve, at $\Omega = 0.6$ is shown in Fig. 12(b). The main directions of the elastic wave propagation are along the ±$y$-directions. And this directional wave propagation can exist in a frequency bandwidth near $\Omega = 0.6$.

To verify the above prediction, we calculate the transient response of the structure to a harmonic (sinusoidal) wave. The finite structure is composed of $21 \times 21$ unit cells. A point source is located at the center of the central unit cell of the structure. That is to say, a harmonic displacement field with unit amplitude is imposed on the central node of the central unit cell. It operates at a reduced frequency of $\Omega = 0.6$. Here, both *horizontal* (1, 0) [Fig. 13(a)] and *vertical* (0, 1) [Fig. 13(b)] point sources are considered [30]. Their deformed displacement fields at different times are shown in Fig. 13. For elastic waves with either horizontal or vertical polarization, the wave propagation is confirmed mainly within two branches along the ±$y$-direction, which validates the prediction that follows in Fig. 12. It is also noted that the amplitude of the displacement field of the $x$-polarized wave is smaller that of the $y$-polarized wave. To explain this, we calculate the amount of polarization [Eq. (3)] of the cross point A in Fig. 5(b). The value along the $x$-direction is 0.03, which is smaller than that along the $y$-direction. So this is a dominantly $y$-polarized wave, and the radiation of the $y$-polarized source is preferred. The transmission spectra in Fig. 5(b) also illustrate this phenomenon. The transmission coefficient of the $x$-polarized wave (the solid line) is smaller than that of the $y$-polarized wave (the dashed line) at $\Omega = 0.6$.

The harmonic response of the finite symmetric 1d-zigzag lattice is also calculated and presented in Fig. 14. The calculation model consists of a system with $21 \times 21$ unit cells surrounded by the substrate material with the same elastic properties. PMLs are used outside of the substrate material to avoid the wave reflection at the fictitious boundaries. The directional properties are also shown clearly in the deformed displacement fields in Figs. 14(a) and 14(b). A good agreement between the transient response and the frequency response is observed. The slight difference of the wave propagation in +*y*- and –*y*-directions are owing to the symmetry of the system.

Next, we examine the directional wave propagation of the antisymmetric 2d-zigzag lattice. Isofrequency curves of the 7th and 8th phase constant surfaces, associated with the band structures for $\theta = 60^\circ$ in Fig. 8(a), are illustrated in Figs. 15(a) and 15(b). The curves are nearly parallel to the axes. To clearly show the directional propagation of the elastic wave, we present the direction of the group velocity at $\Omega = 0.68$ in the polar plot in Fig. 15(c) which shows that the main directions of the elastic wave propagation slightly deviate from the *x*- and *y*-directions by a small angle of about $3^\circ$. The corresponding results for the 9th phase constant surface are presented in Figs. 15(d) and 15(e). The slopes of the curves are higher than those of the 7th and 8th phase constant surfaces. So the main directions of the wave propagation at $\Omega = 0.84$ are near ±45° and far from the bending arm directions.

We further calculate the transient response of a finite 2d-zigzag lattice with $21 \times 21$ unit cells subjected to a harmonic load. The *horizontal* source (1, 0) is located at the center of the structure. The deformed displacement fields at different times are turgidly shown in Fig. 16(a) ($\Omega = 0.68$) to highlight the directionality. The results show that the directions of the elastic wave propagation are in general agreement with the evaluation predicted by Fig. 14. However, one may also notice that the deformation is mainly localized along the bending arm direction in Fig. 16(a), while in Fig. 16(b) ($\Omega = 0.84$) this effect is less pronounced. So the concentricity for the wave propagation along the bending arm direction [Fig. 16(a)] is better than that along the diagonal [Fig. 16(b)]. Besides, it is noted that the directional wave propagation in Fig. 16(a) is not precisely along the diagonal, but with a small deviation about 3°. So the wave propagates along the +48° direction other than the −42° direction for the horizontally exciting source (1, 0). If a vertically exciting source (0, 1) is applied, the wave will mainly propagate along the −42° direction.

It is also noted that the *x*-direction is preferred to some extent in Fig. 16(a) because the source is *x*-polarized. The coexistence of the longitudinal and transverse wave modes results in this different phenomenon compared to the flexural wave [20, 21] or the acoustic wave [31, 32]. Actually, the *x*- and *y*-directions are identical if the sources have the same symmetry as the system. To show this, we consider a group of four point sources which are located at $(\pm a/2, 0)$ and $(0, \pm a/2)$ with the amplitudes being $(\pm 1, 0)$ and $(0, \pm 1)$, respectively. Here, '−1' means that the vibration is in the reverse phase of the one with '+1'. This particular setting guarantees the same symmetry of the sources and the structures. The deformed displacement fields at different times are shown in Fig. 17. Unlike the case with one point source in Fig. 16(a), the wave propagation contains mainly four identical branches as expected.

### 5.2 The perforated 2d-zigzag lattice with cross holes

Next we consider the directional wave propagation in the perforated 2d-zigzag lattice with cross holes. The isofrequency curves of the 6th, 7th and 5th phase constant surfaces, associated with the band structures for $\varphi = 30^\circ$ in Fig. 10, are shown in Figs. 18(a), 18(b) and 18(d), respectively. The directional wave propagation at $\Omega = 0.44$ and $\Omega = 0.3$ in the polar plot are also presented in Fig. 18. It is shown in Fig. 18(c) that the main directions of the elastic wave propagation at $\Omega = 0.44$ are near ±*x* and ±*y*-directions. While in Fig. 18(e), the directions at $\Omega = 0.3$ are close to ±45°.

The transient responses of a finite structure with $21 \times 21$ unit cells to harmonic loads are also calculated, and the deformed displacement fields at different times are presented in Figs. 19 and 20. Both a single point source (Fig. 19) and a group of four point sources (Fig. 20) are considered. Different patterns of directional wave propagation are observed. In Fig. 18(e), the main direction of the wave propagation has a small deviation (−0.23°) from ±45°, so the wave propagation along the two diagonals is not identical in Fig. 19(b). The concentricity of the directional wave propagation along the bending arm direction [Fig. 19(a)] is much more pronounced than that along the diagonals [Fig. 19(b)]. When the symmetry of the point sources is the same as that of the structure, the four branches of the directional wave propagation are identical, as shown in Fig. 20. These results are similar to those of the antisymmetric 2d-zigzag lattice as shown in Figs. 16 and 17.

Actually, a system with straight arms can also exhibit directional wave propagation. To compare the directional properties of the proposed systems with the one having straight arms, we consider the directional wave propagation along the *x*-axis. A concentration degree of the directional wave is defined as follows. We consider the displacement distribution along the *y*-axis for a fixed *x*, which is normalized by its maximal value, and then calculate the inertia moment of the normalized displacement distribution and average it over the length of the structure in the *x*-direction, which is given as

$$I_x^t = \frac{1}{l_x}\int \frac{1}{l_y}\int \left[ \frac{\sqrt{|\mathrm{u}|^2+|\mathrm{v}|^2}}{(\sqrt{|\mathrm{u}|^2+|\mathrm{v}|^2})_{\max}} |y-y_0|^2 \right] \mathrm{d}y\mathrm{d}x, \tag{5}$$

where $l_x$ and $l_y$ are the lengths of the finite structure in *x*- and *y*- directions, and $y_0$ is the location of the excitation. The value obtained from Eq. (5) is averaged once more over a number of time period, i.e.

$$\bar{I}_x = \sqrt{\frac{1}{NT}\int_0^{NT} (I_x^t)^2 \mathrm{d}t}, \tag{6}$$

where *N* is the number of the period *T*. Finally, the concentration degree is defined as

$$\bar{C}_x = 1/\bar{I}_x = 1\Big/\sqrt{\frac{1}{NT}\int_0^{NT} (c_x^t)^2 \mathrm{d}t}. \tag{7}$$

The bigger the concentration degree is, the better the concentricity of the directional wave propagation is. Here, *N*=2 is chosen to avoid the wave reflections at the fictitious boundaries. The concentration degree is obtained as $\bar{C}_x = 5.6\times10^6\,\mathrm{m}^{-2}$ for the 2d-zigzag lattice in Fig. 16(a) and $\bar{C}_x = 1.4\times10^7\,\mathrm{m}^{-2}$ for the system with rotated cross holes in Fig. 19(a). While for the system with straight arms shown in Fig. 5(a), the average concentration degree is $\bar{C}_x = 2.5\times10^6\,\mathrm{m}^{-2}$ at $\Omega = 0.66$, which is rather smaller than those of the zigzag lattices. So, directional wave propagation in the zigzag lattices is more concentrated.

## 6. Conclusion

In this paper, bandgap and directional behaviors of elastic wave propagation in zigzag lattice structures are analyzed by using the finite element method. Two types of structures with the bending arms in one or two directions (termed as 1d- and 2d-zigzag lattices, respectively) are considered. The transmission spectra of the systems are calculated. The effects of the geometry parameters of the 2d-zigzag lattices on the bandgaps are investigated and discussed. The mechanism of the bandgap generation is analyzed by studying the vibration modes at the bandgap edges. Directional propagation of elastic waves in the considered systems are also investigated. The deformed displacement fields of the dynamic response of different finite systems to different harmonic loads (according to the symmetry of the system) are presented. From the obtained results and discussions we can draw the following conclusions:

1) No complete bandgap appears in a two-dimensional square lattice structure with only straight arms. By introducing zigzag lattices, the degeneracy of some bands is separated due to the complex vibration modes (especially the rotational modes) of the bending arms. Different types of the symmetry can lead to different vibration modes and give rise to different directional or complete bandgaps. Multiple wide complete bandgaps can be found in a wide porosity range. Distinguished features can be obtained in various kinds of zigzag lattices. Generally, multiple directional bandgaps are generated in 1d-zigzag lattices, and multiple wide complete bandgaps are obtained in 2d-zigzag lattice.

2) Detailed study of the geometry parameters for the 2d-zigzag lattice shows that an intermediate value of the bending angle of the arms or the rotational angle of the cross holes is favorable for the generation of a wide bandgap. Therefore, wider and lower bandgaps may be obtained by optimizing the geometry of the zigzag lattice structures.

3) Directional propagation of elastic waves in different directions and at different frequencies can be realized by a careful design of zigzag lattice structures. The directional wave propagation in the proposed structures is more concentrated than that in the system with straight arms. Generally, the concentricity of the directional wave propagation is perfect when the energy propagates along the straight arm direction in 1d-zigzag lattice or along the bending arm direction in the 2d-zigzag lattice. The proposed structures are potential candidates of acoustic/elastic metamaterials.

## Acknowledgements

The first two authors are grateful to the support from the National Natural Science Foundation of China (11272041) and the National Basic Research Program of China (2010CB732104). The first and the third authors acknowledge the financial support by the German Research Foundation (DFG, No. ZH 15/16-1) and International Bureau of the German Federal Ministry of Education and Research (BMBF, No. CHN 11/045). The first author also acknowledges the Visiting Fund of Beijing Jiaotong University for the research stay at the Chair of Structural Mechanics, Department of Civil Engineering, University of Siegen, Germany.

# Figure captions

Fig. 1 Sketch of two-dimensional periodic structures, unit cells and their corresponding irreducible Brillouin zones for antisymmetric (a) 1d- and (b) 2d-zigzag lattices.

Fig. 2 Unit cells and their corresponding irreducible Brillouin zone for symmetric (a) 1d- and (b) 2d-zigzag lattices.

Fig. 3 (a) An alternative profile of 2d-zigzag lattice and (b) its unit cell.

Fig. 4 Models for calculating the transmission coefficient of systems along the (a) $\Gamma X$ and (b) $\Gamma M$ directions.

Fig. 5 Band structures of (a) square lattice with straight arms, (b) antisymmetric and (c) symmetric 1d-zigzag lattices, and (d) antisymmetric and (e) symmetric 2d-zigzag lattices. Their porosities are the same (*f*=0.8). Transmission spectra of the zigzag lattices are also presented. The solid and dashed lines represent the *x*- and *y*-polarized waves, respectively. The light gray represents the directional bandgaps in the square lattice with straight arms in panel (a) and those in the other systems between the same bands, and the dark gray represents the new directional bandgaps appearing in the zigzag lattices. The diagonal lines represent the complete bandgaps. The separation of the degeneracy at the corresponding crossover regions in panel (a) is demonstrated by the arrows.

Fig. 6 Vibration modes at the points marked in Fig. 5.

Fig. 7 (a) Band structures for cross holes with $\varphi = 30^{\circ}$, (b) the vibration mode at point $X_4^3$ in panel (a). The dashed line shown in panel (a) represents the band structure of zigzag lattice with straight arms in Fig. 5(a).

Fig. 8 (a) Band structures of antisymmetric 2d-zigzag lattices with various bending angles. For each angle, the horizontal axis represents the wave vector via $\Gamma$-X-M-$\Gamma$. Panels (b) and (c) present the vibration modes at the points marked in panel (a).

Fig. 9 Variations of the bandgap edges for antisymmetric 2d-zigzag lattices with (a) different bending angles but the same porosity and bending distance, (b) different porosities but the same arm width and bending distance, (c) different porosities but the same bending angle and bending distance, and (d) different bending distances but the same porosity and bending angle.

Fig. 10 (a) Band structures for cross holes with different rotation angles. For each angle, the horizontal axis represents the wave vector via $\Gamma$-X-M-$\Gamma$. Panels (b) and (c) present the vibration modes at the points marked in panel (a).

Fig. 11 Variations of the bandgap edges with (a) the rotational angle, or the geometry size (b) *b*/*a* and (c) *c*/*a* of the cross hole.

Fig. 12 (a) Isofrequency curves of the 6th phase constant surface associated with the band structure in Fig. 5(c) for symmetric 1d-zigzag lattice. (b) Polar plot of the directional wave propagation at $\Omega = 0.6$, see the dashed line marked in Fig. 5(c).

Fig. 13 The transient response for one point source of (a) horizontal and (b) vertical polarization at $\Omega = 0.6$ for symmetric 1d-zigzag lattice with $T = 2\pi/\omega$.

Fig. 14 The harmonic response for one point source of (a) horizontal and (b) vertical polarization at $\Omega = 0.6$ for symmetric 1d-zigzag lattice.

Fig. 15 Isofrequency curves of (a) the 7th and (b) 8th phase constant surface associated with the band structure in Fig. 8 for antisymmetric 2d-zigzag lattice. (c) Polar plot of the directional wave propagation at $\Omega = 0.68$, see the thick dashed line marked in Fig. 8. (d) Isofrequency curves of the 9th phase constant surface. (e) Polar plot of the directional wave propagation at $\Omega = 0.84$, see the thin dashed line marked in Fig. 8.

Fig. 16 The transient response at (a) $\Omega = 0.68$ and (b) $\Omega = 0.84$ for antisymmetric 2d-zigzag lattice.

Fig. 17 The transient response for a group of four point sources at $\Omega = 0.68$ for antisymmetric 2d-zigzag lattice.

Fig. 18 Isofrequency curves of the (a) 6th and (b) 7th phase constant surface associated with the band structure in Fig. 10 for system with rotated cross holes. (c) Polar plot of the directional wave propagation at $\Omega = 0.44$, see the thick dashed line marked in Fig. 10. (d) Isofrequency curves of the 5th phase constant surface. (e) Polar plot of the directional wave propagation at $\Omega = 0.3$, see the thin dashed line marked in Fig. 10.

Fig. 19 The transient response at (a) $\Omega = 0.44$ and (b) $\Omega = 0.3$ for system with rotated cross holes.

Fig. 20 The transient response for a group of four point sources at $\Omega = 0.44$ for system with rotated cross holes.

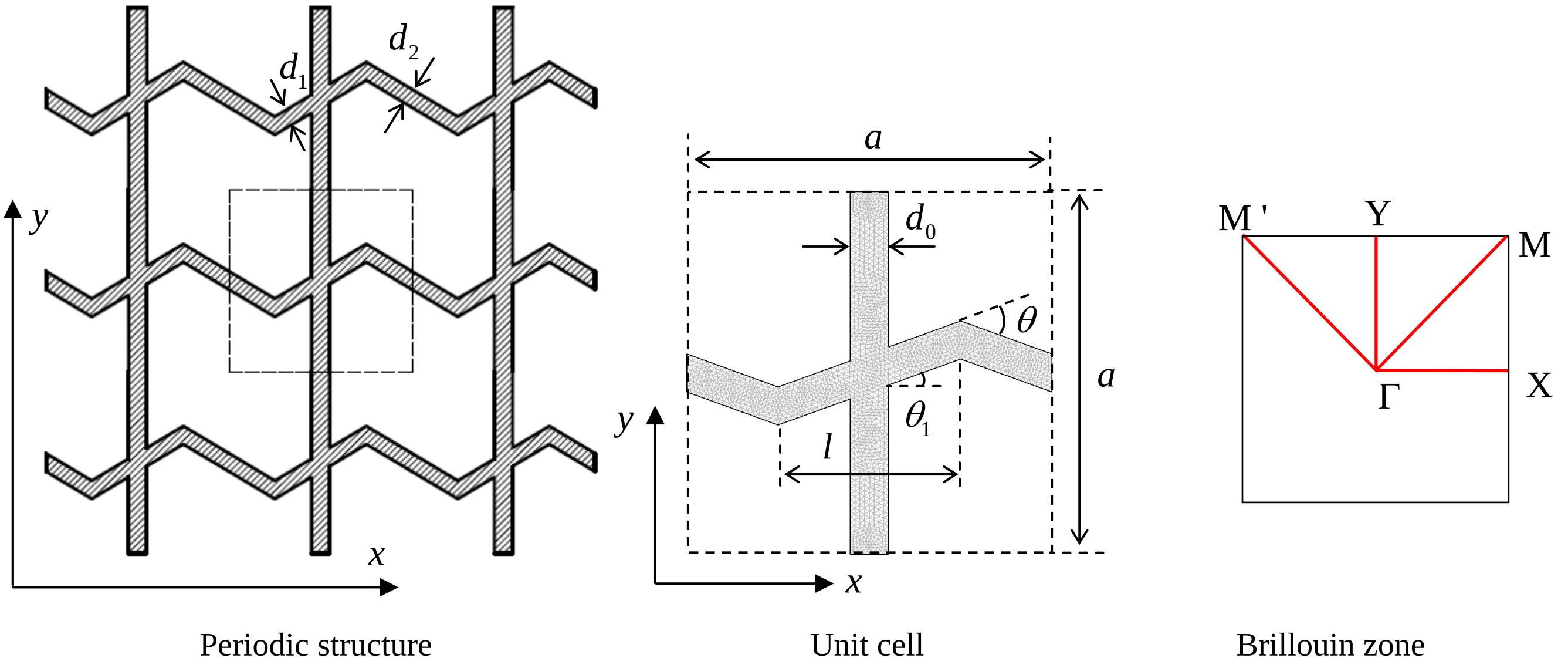

(a)

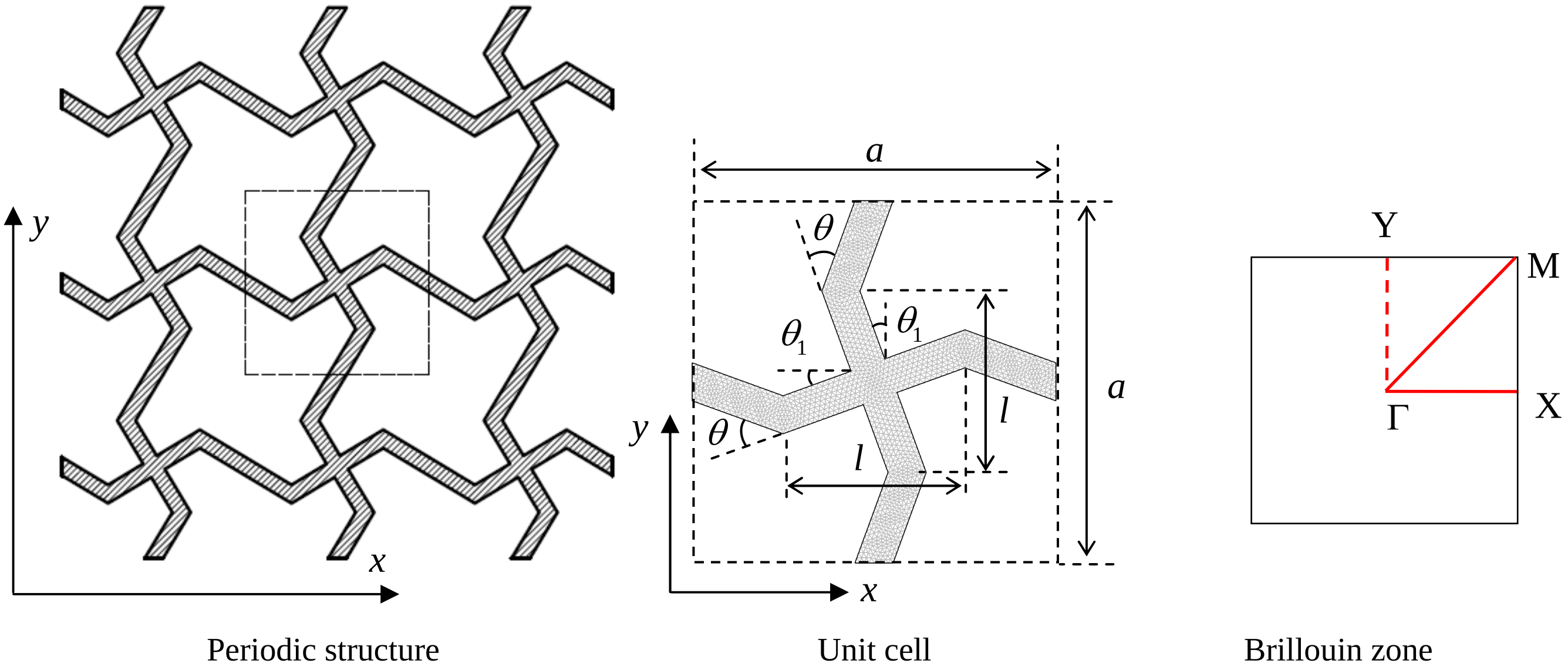

(b)

Fig. 1 Sketch of two-dimensional periodic structures, unit cells and their corresponding irreducible Brillouin zones for antisymmetric (a) 1d- and (b) 2d-zigzag lattices.

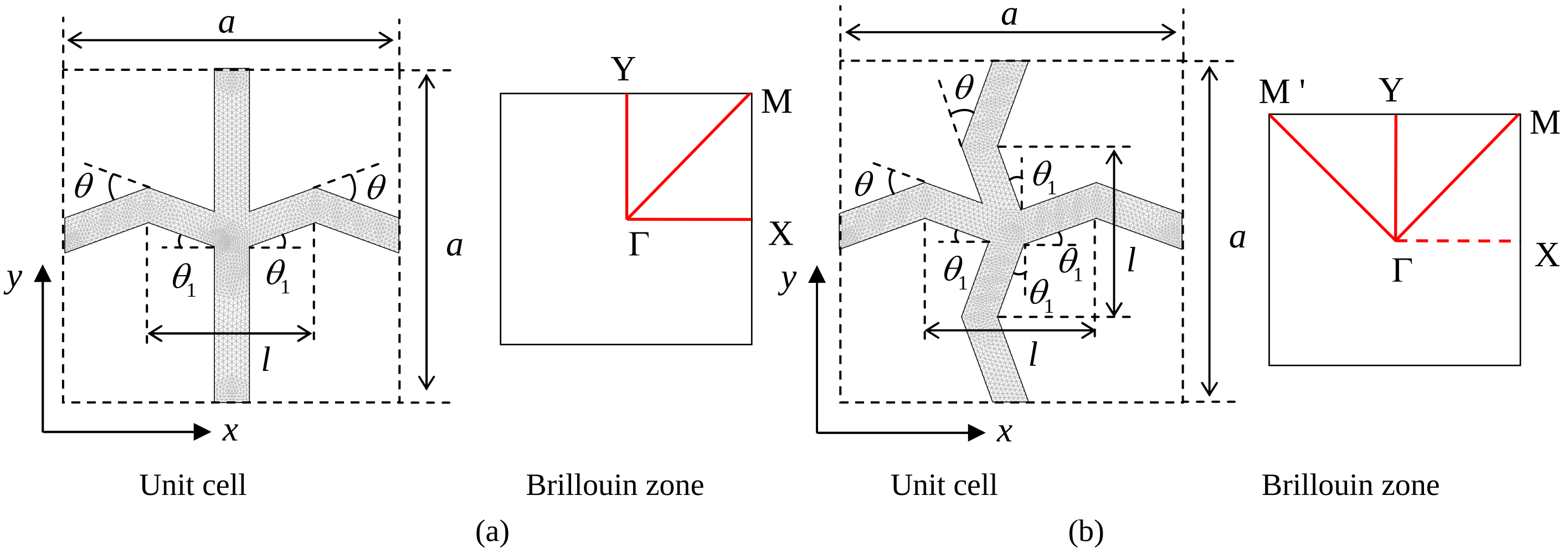

Fig. 2 Unit cells and their corresponding irreducible Brillouin zone for symmetric (a) 1d- and (b) 2d-zigzag lattices.

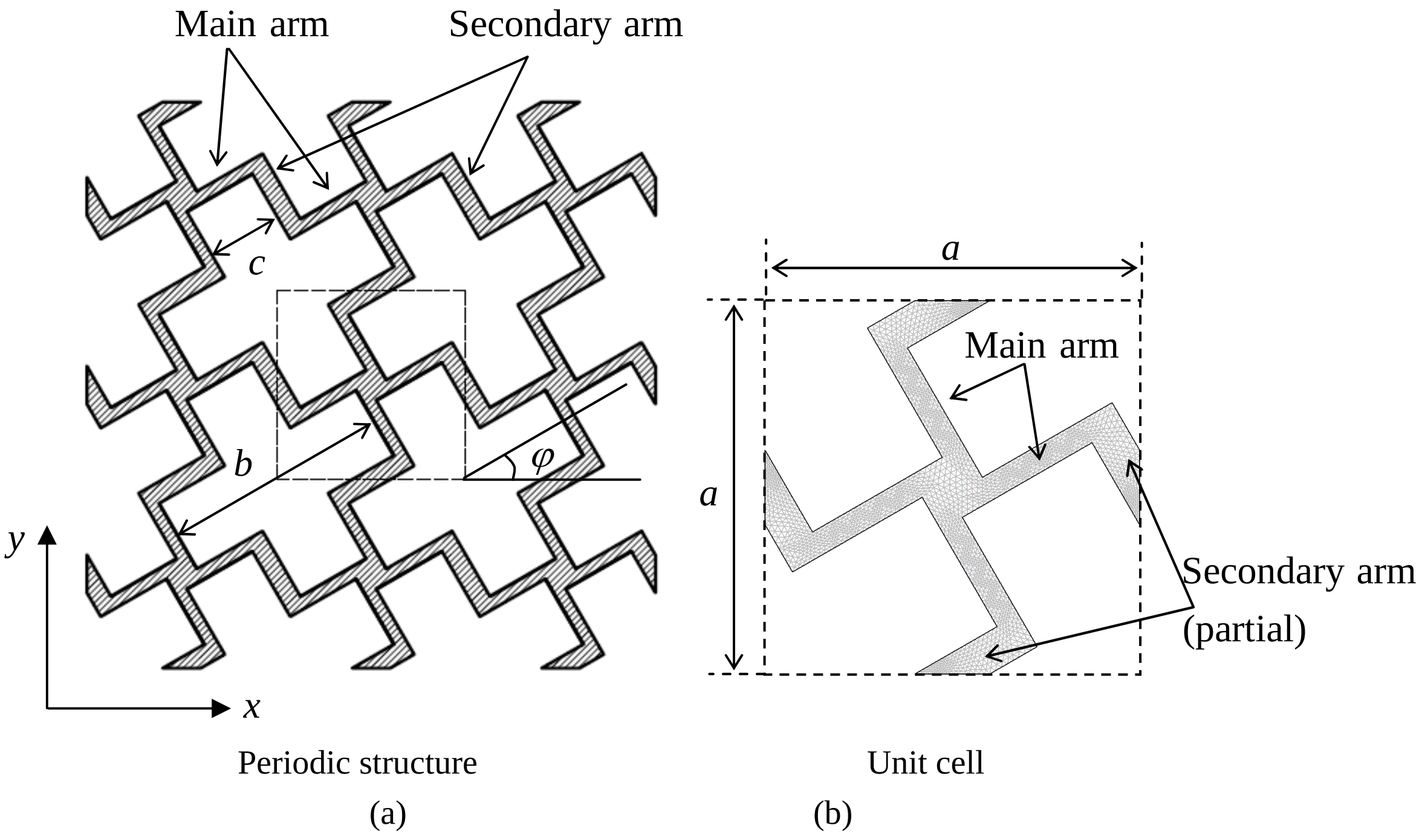

Fig. 3 (a) An alternative profile of 2d-zigzag lattice and (b) its unit cell.

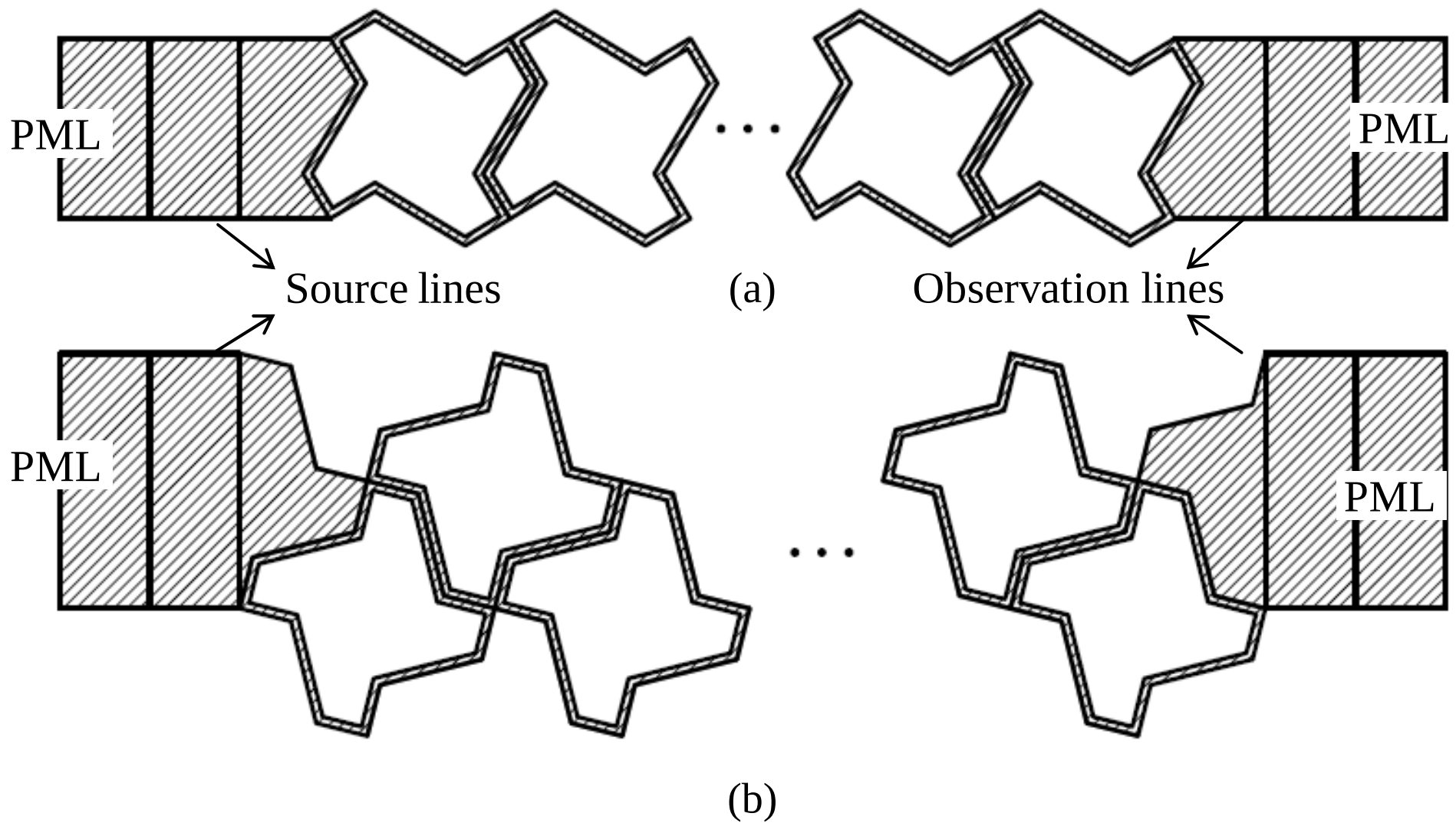

Fig. 4 Models for calculating the transmission coefficient of periodic systems along the (a) ΓX and (b) ΓM directions.

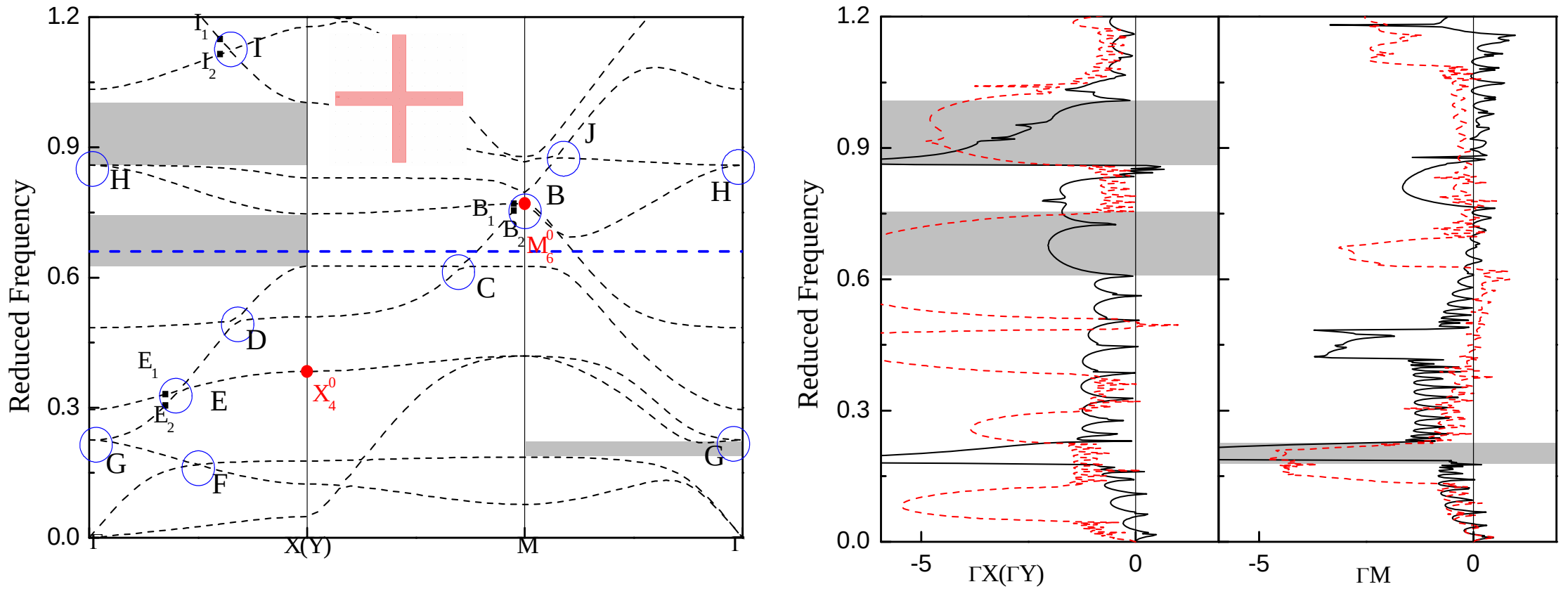

(a)

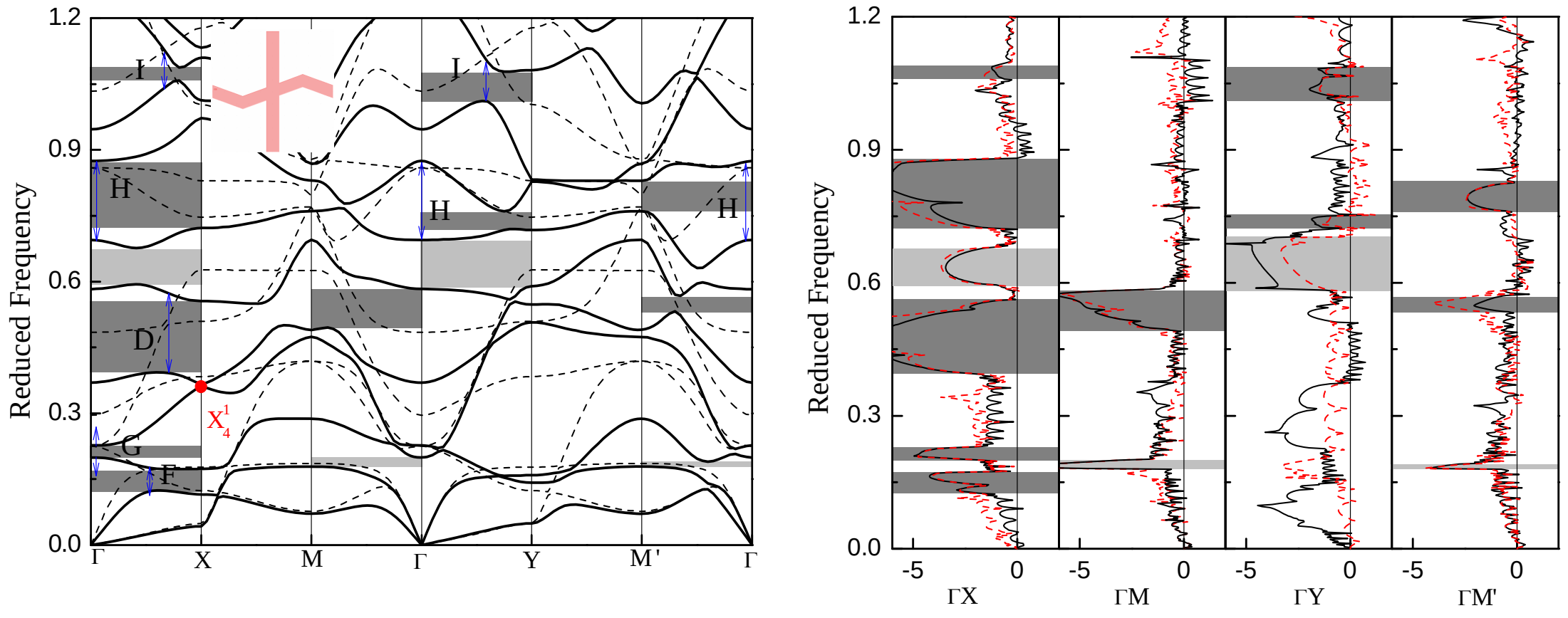

(b)

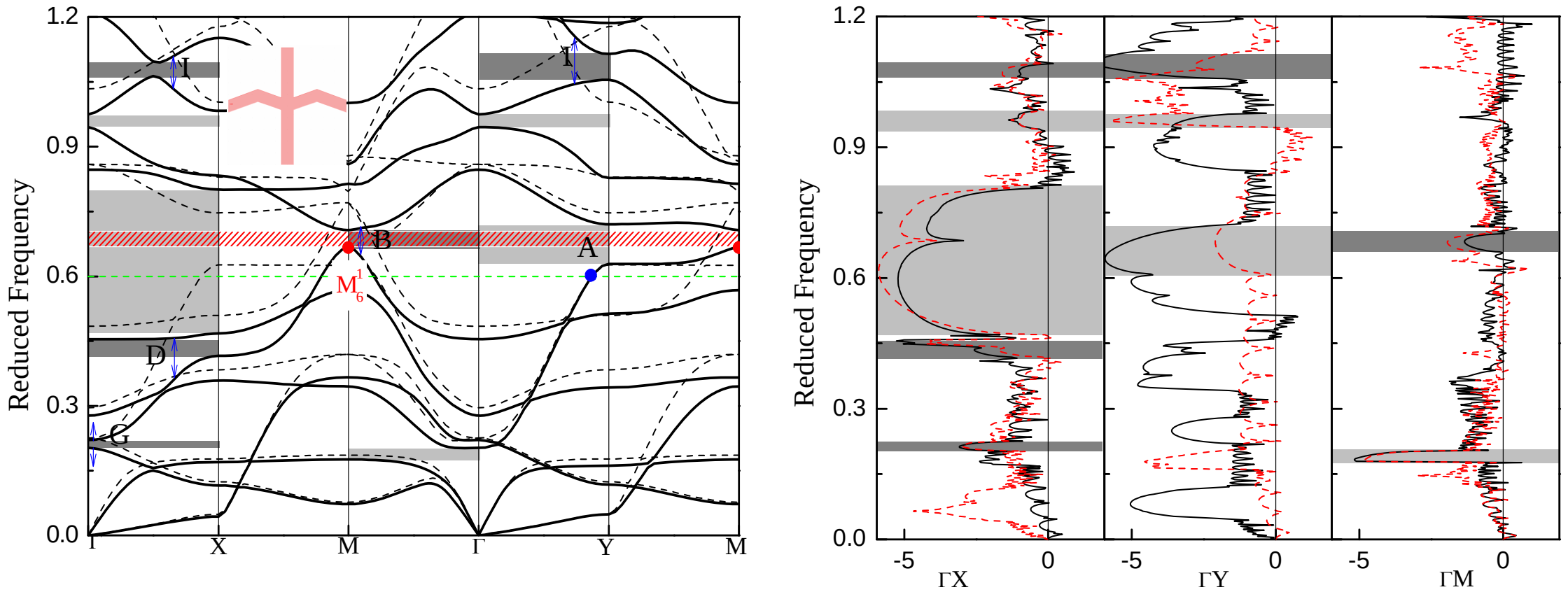

(c)

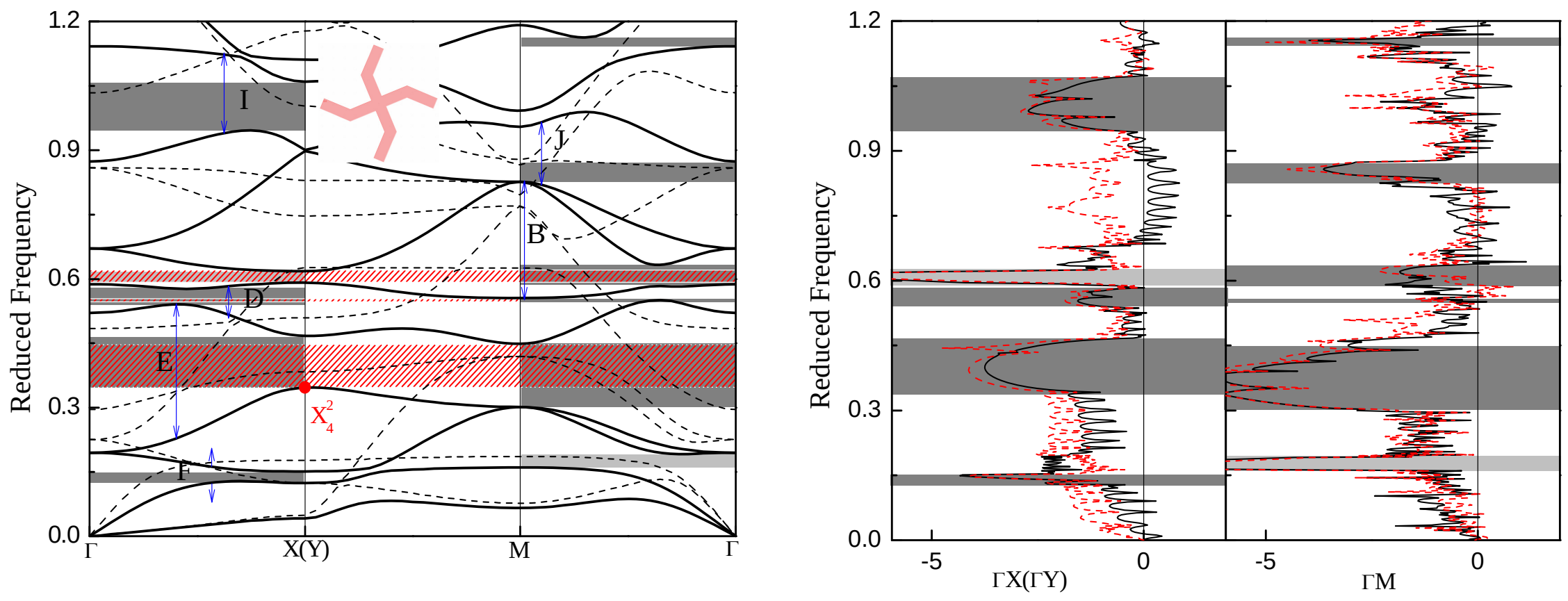

(d)

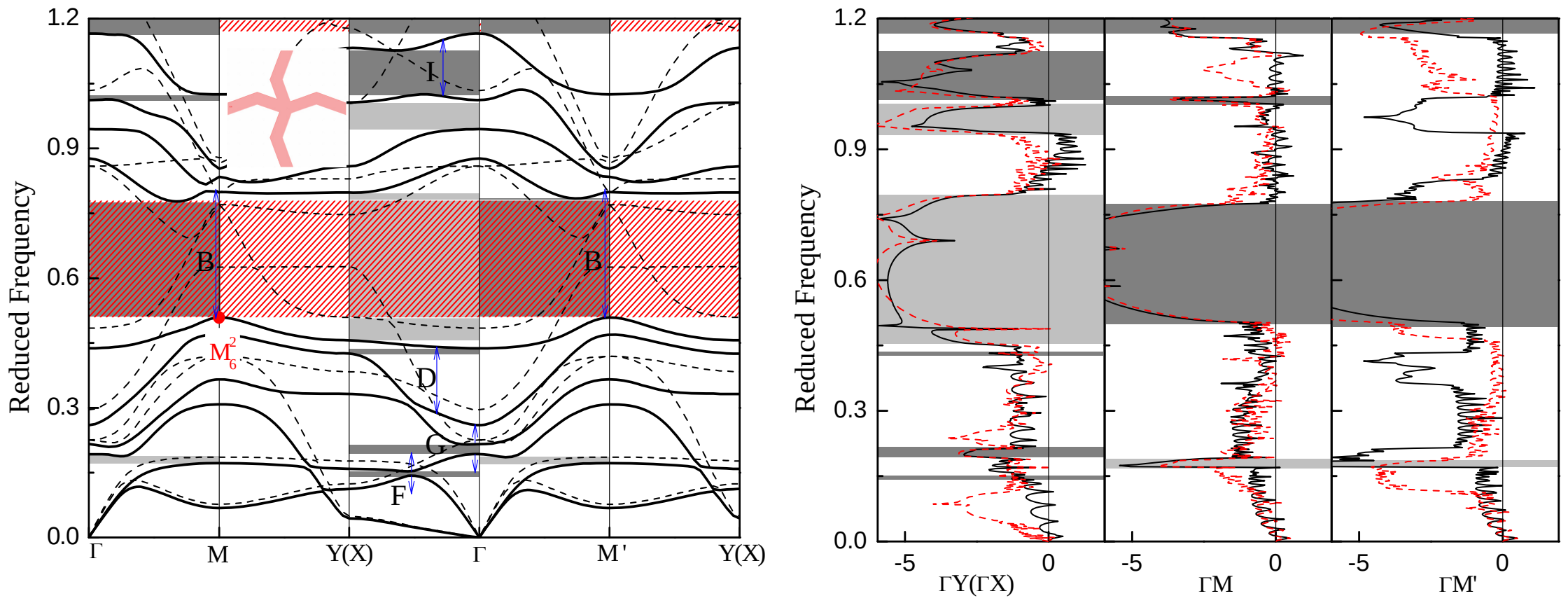

(e)

Fig. 5 Band structures of (a) square lattice with straight arms, (b) antisymmetric and (c) symmetric 1d-zigzag lattices, and (d) antisymmetric and (e) symmetric 2d-zigzag lattices. Their porosities are the same ($f$=0.8). The dashed line shown in panels (b)-(e) represents the band structures in panel (a). Transmission spectra of the zigzag lattices are also presented. The solid and dashed lines represent the $x$- and $y$-polarized waves, respectively. The light gray represents the directional bandgaps in the square lattice with straight arms in panel (a) and those in the other systems between the same bands, and the dark gray represents the new directional bandgaps appearing in the zigzag lattices. The diagonal lines represent the complete bandgaps. The separation of the degeneracy at the corresponding crossover regions in panel (a) is demonstrated by the arrows.

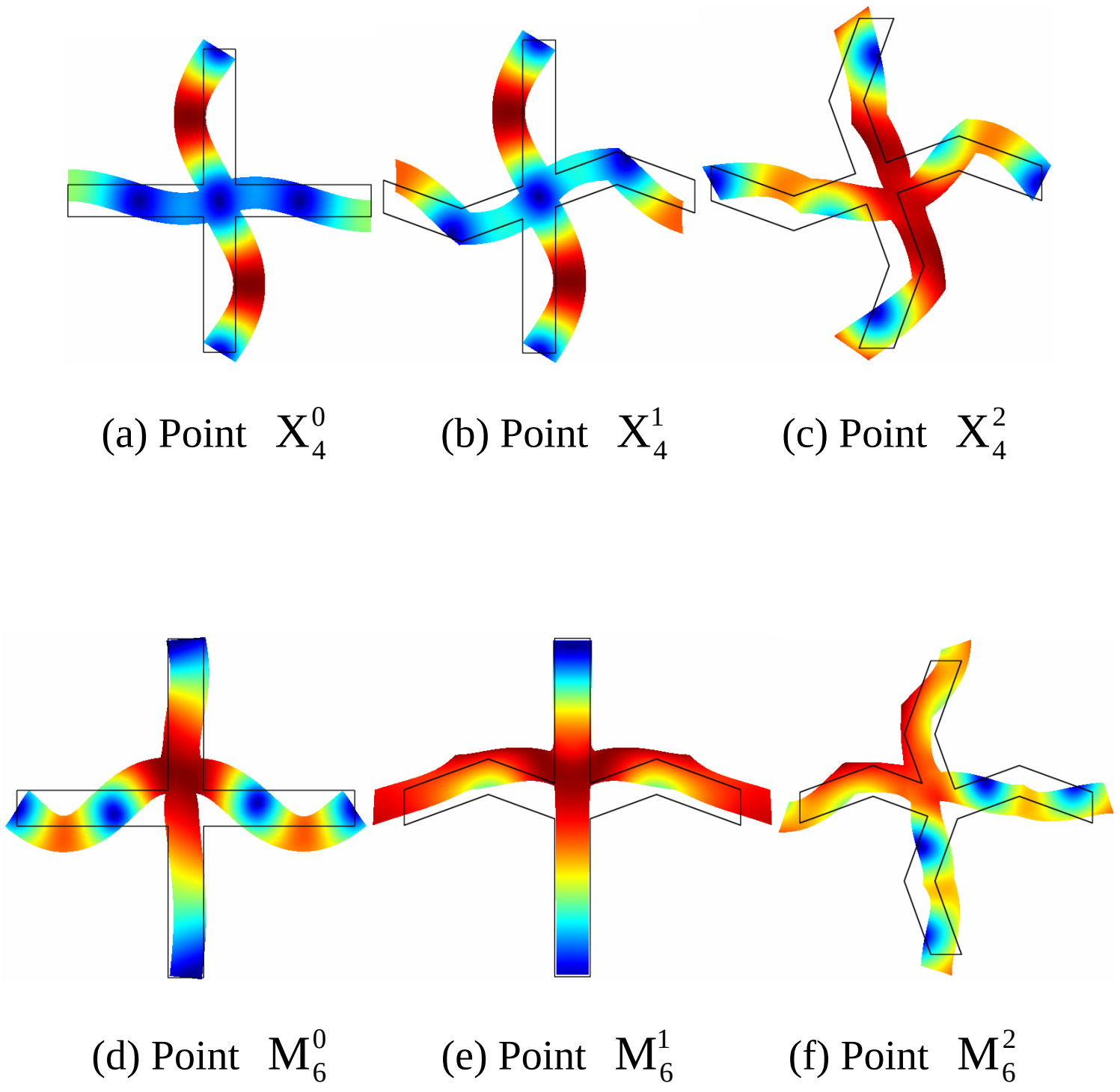

Fig. 6 Vibration modes at the points marked in Fig. 5.

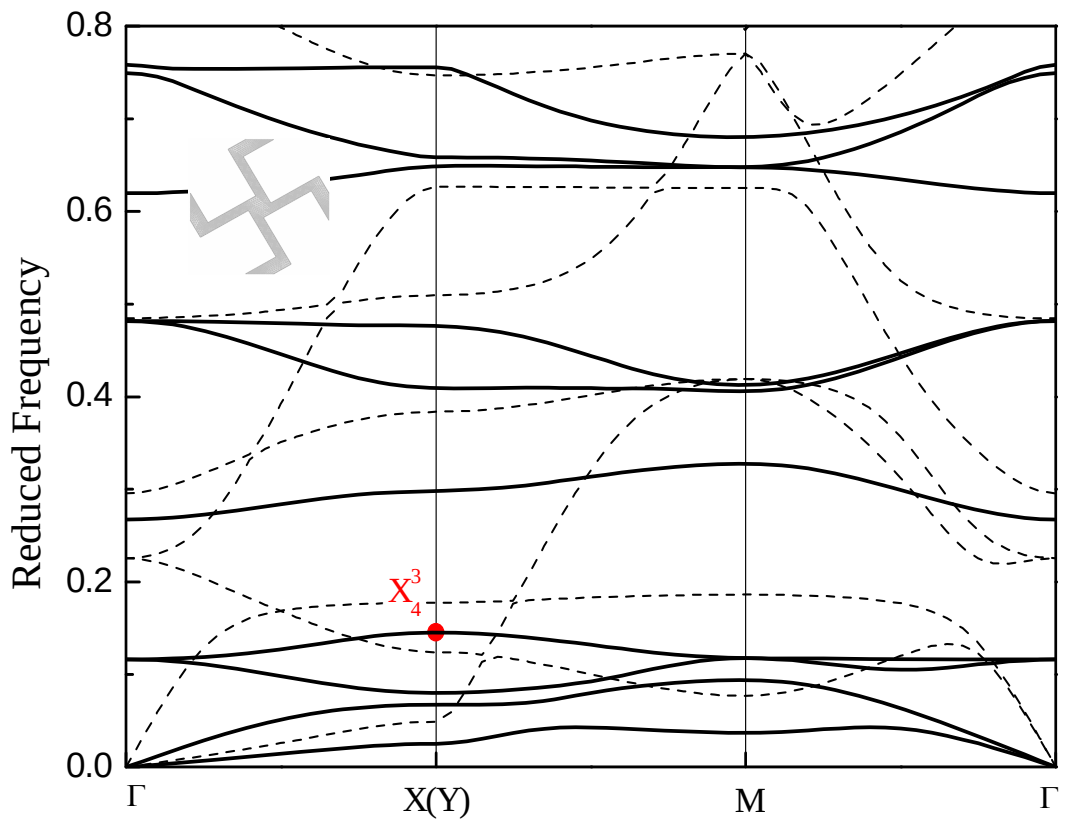

(a)

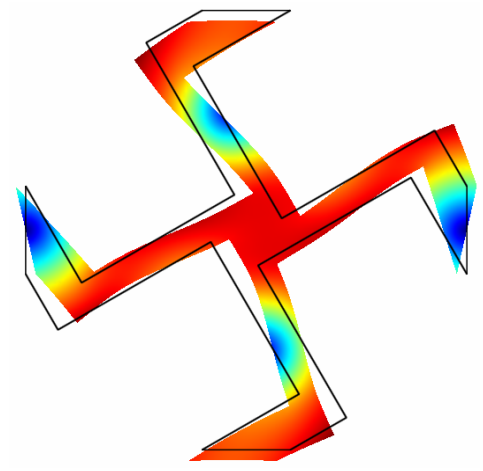

(b)

Fig. 7 (a) Band structures for cross holes with $\varphi = 30^{\circ}$, (b) the vibration mode at point $X_4^3$ in panel (a). The dashed line shown in panel (a) represents the band structure of zigzag lattice with straight arms in Fig. 5(a).

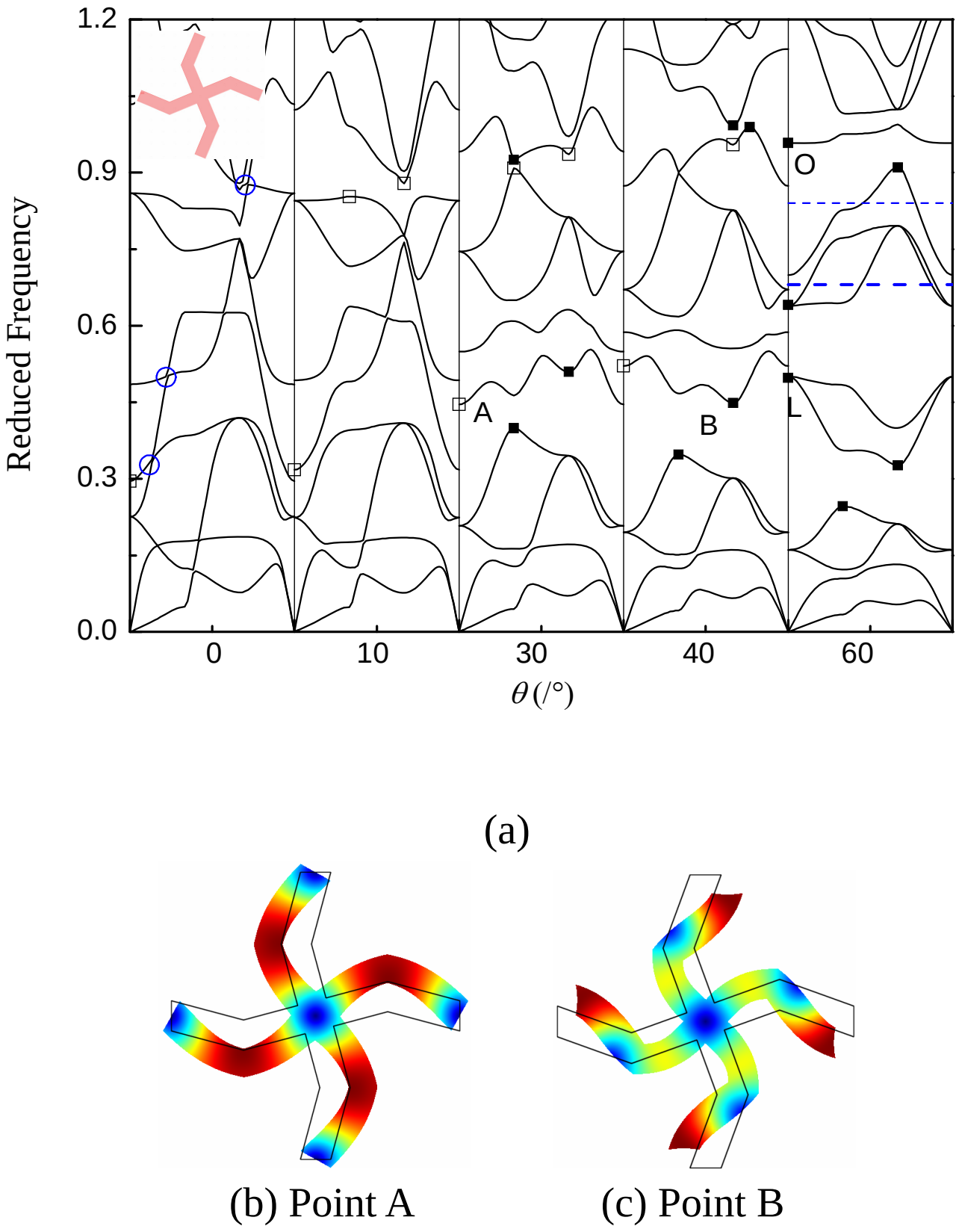

Fig. 8 (a) Band structures of antisymmetric 2d-zigzag lattices with various bending angles. For each angle, the horizontal axis represents the wave vector via Γ-X-M-Γ. Panels (b) and (c) present the vibration modes at the points marked in panel (a).

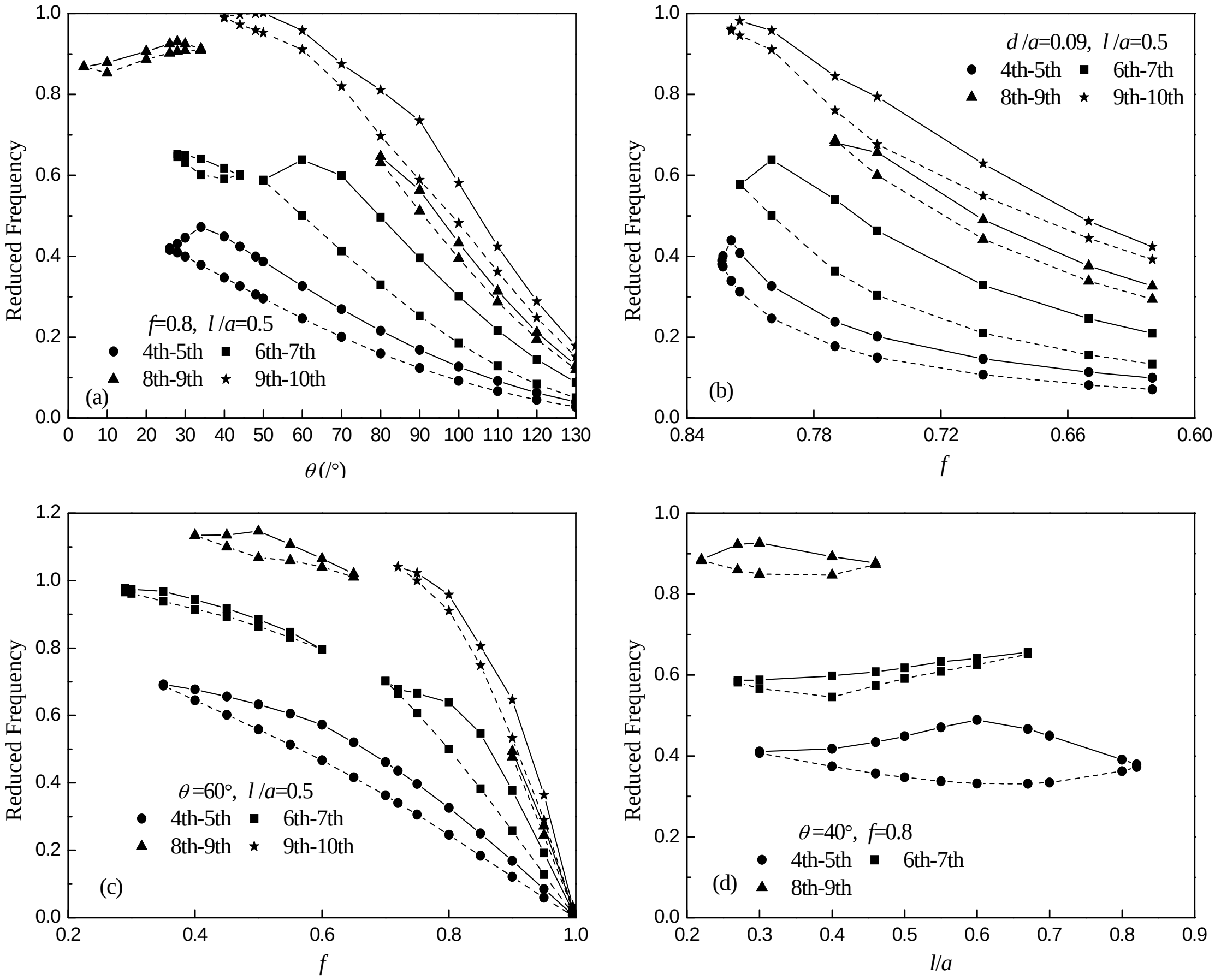

Fig. 9 Variations of the bandgap edges for antisymmetric 2d-zigzag lattices with (a) different bending angles but the same porosity and bending distance, (b) different porosities but the same arm width and bending distance, (c) different porosities but the same bending angle and bending distance, and (d) different bending distances but the same porosity and bending angle.

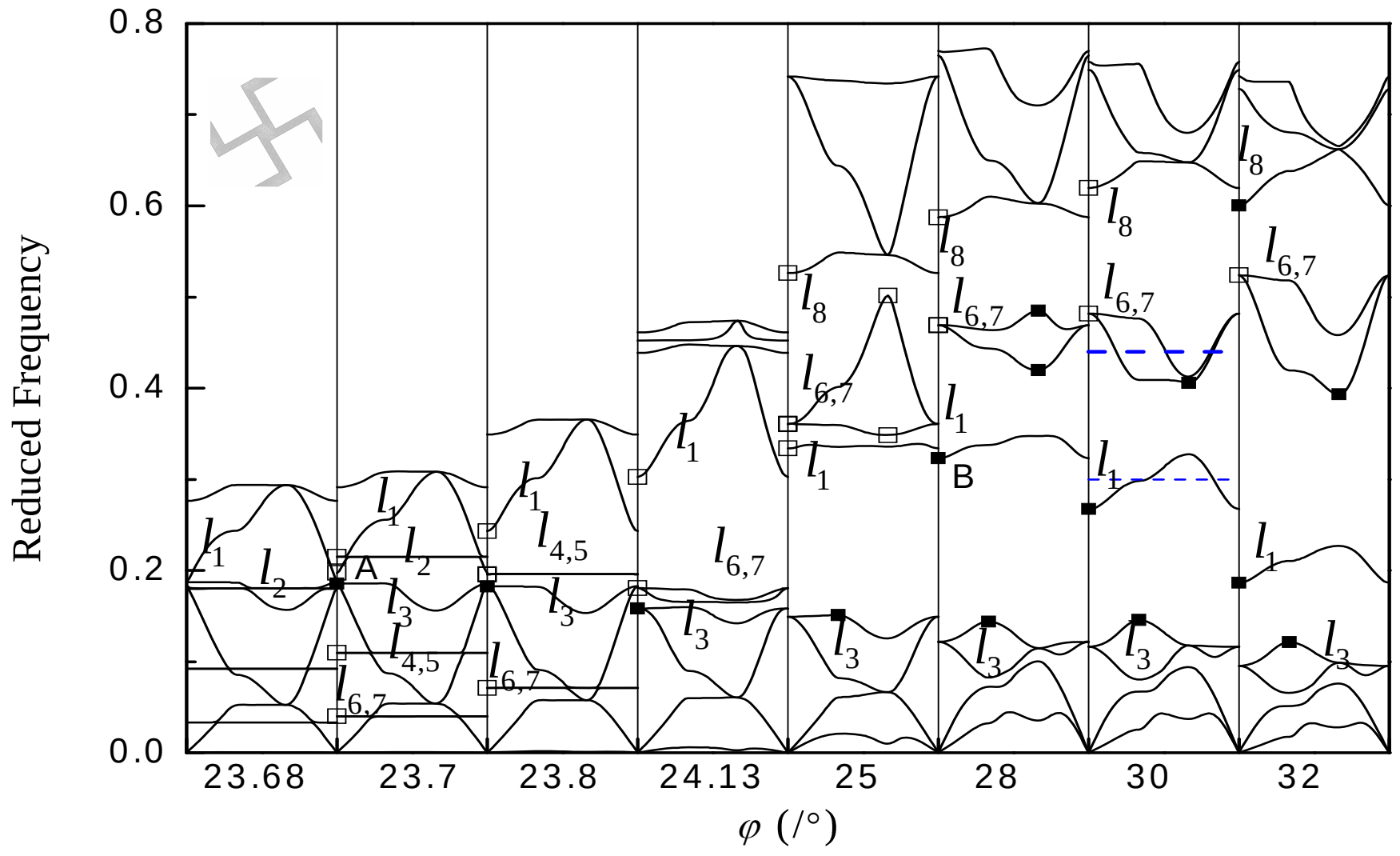

(a)

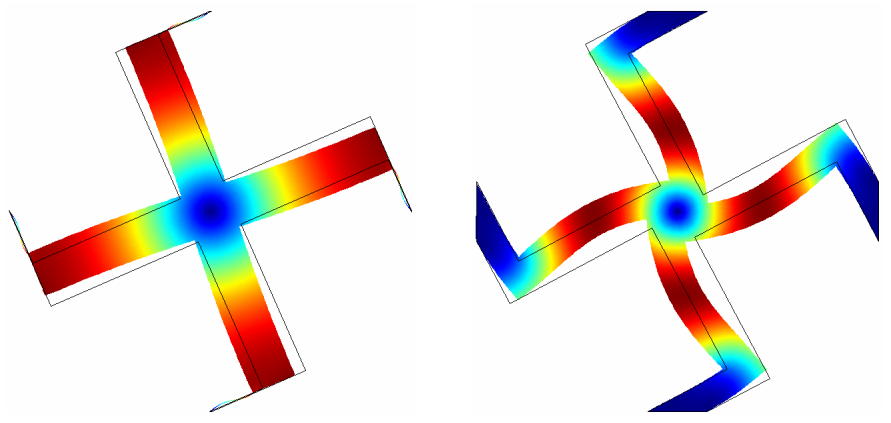

(b) Point A (c) Point B

Fig. 10 (a) Band structures for cross holes with different rotation angles. For each angle, the horizontal axis represents the wave vector via Γ-X-M-Γ. Panels (b) and (c) present the vibration modes at the points marked in panel (a).

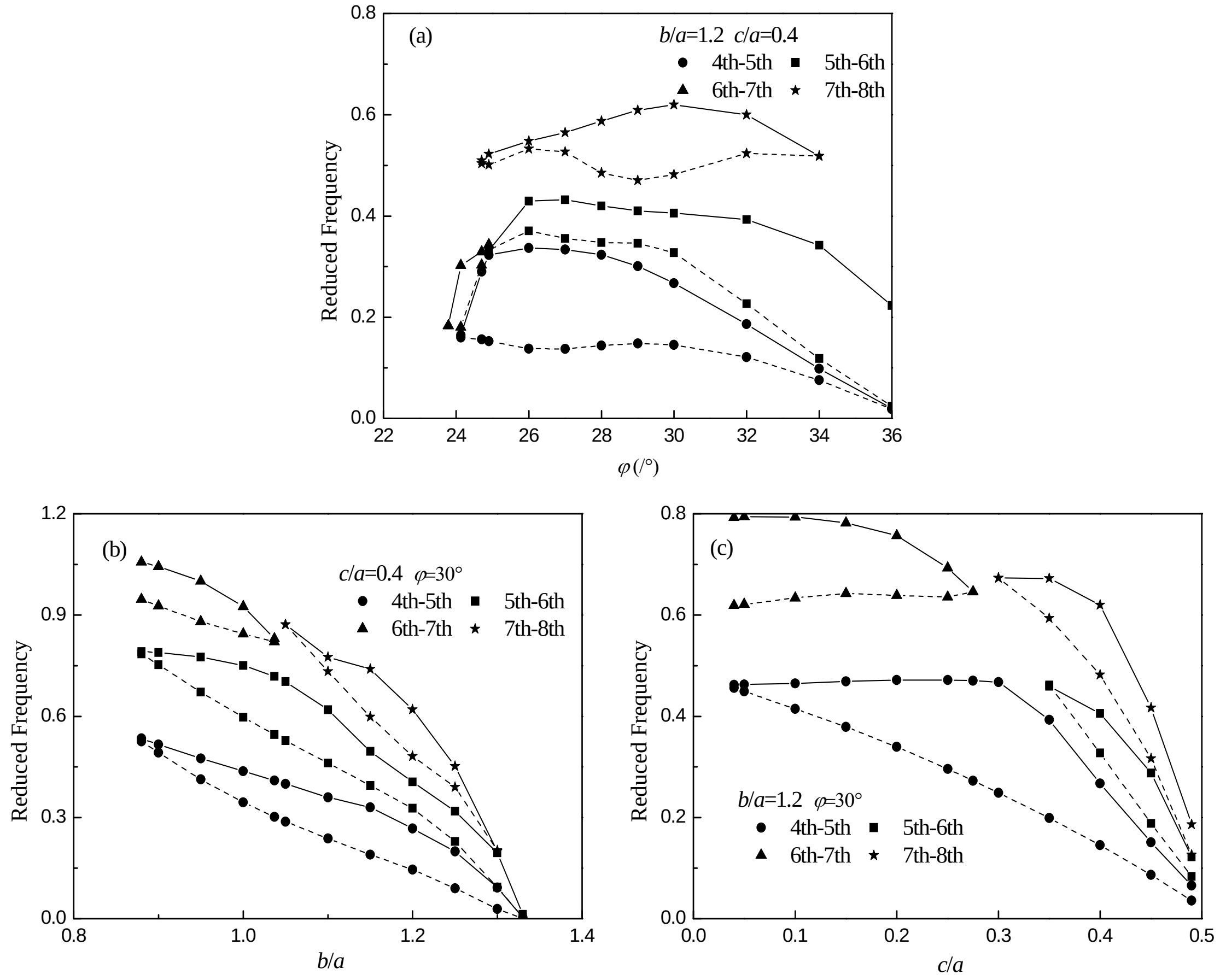

Fig. 11 Variations of the bandgap edges with (a) the rotational angle, or the geometry size (b) *b*/*a* and (c) *c*/*a* of the cross hole.

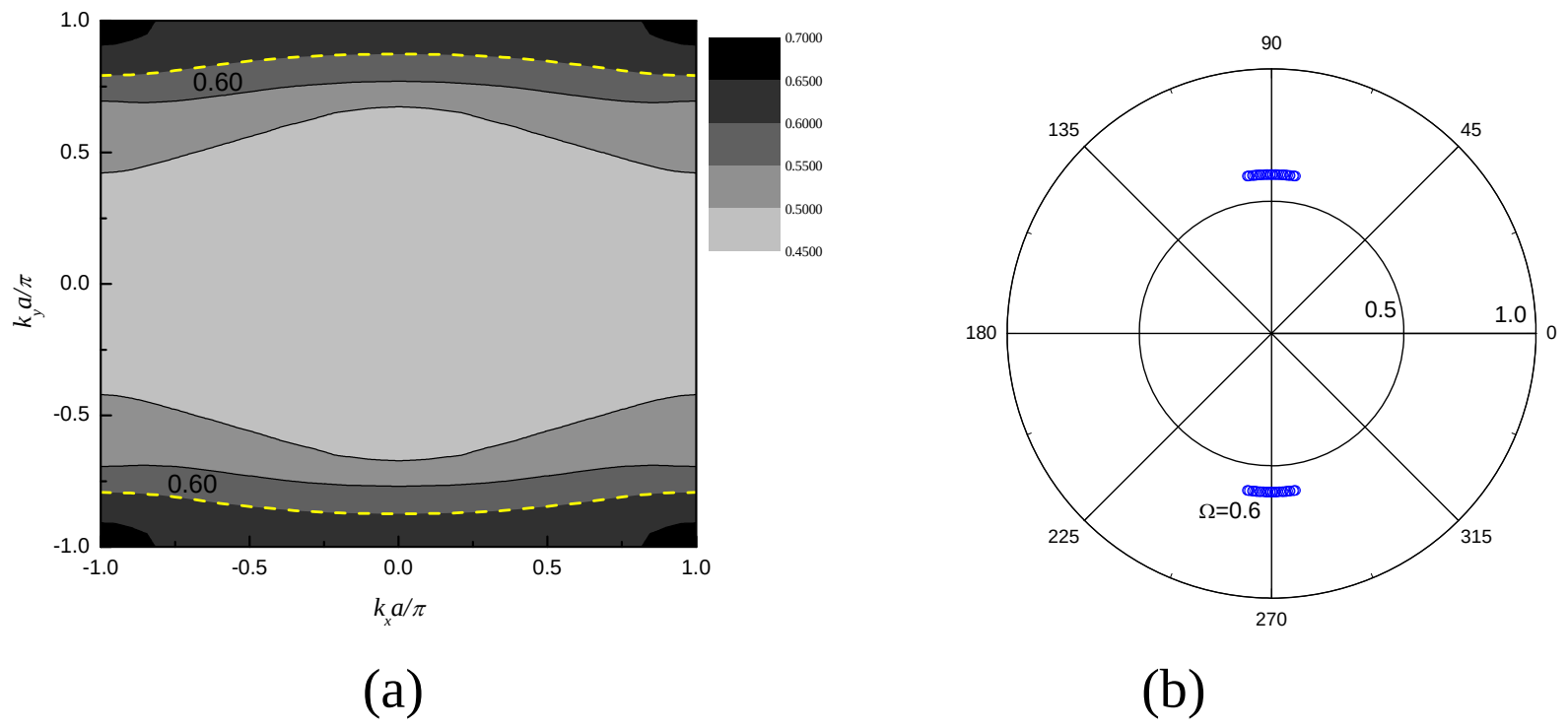

Fig. 12 (a) Isofrequency curves of the 6th phase constant surface associated with the band structure in Fig. 5(c) for symmetric 1d-zigzag lattice. (b) Polar plot of the directional wave propagation at $\Omega = 0.6$, see the dashed line marked in Fig. 5(c).

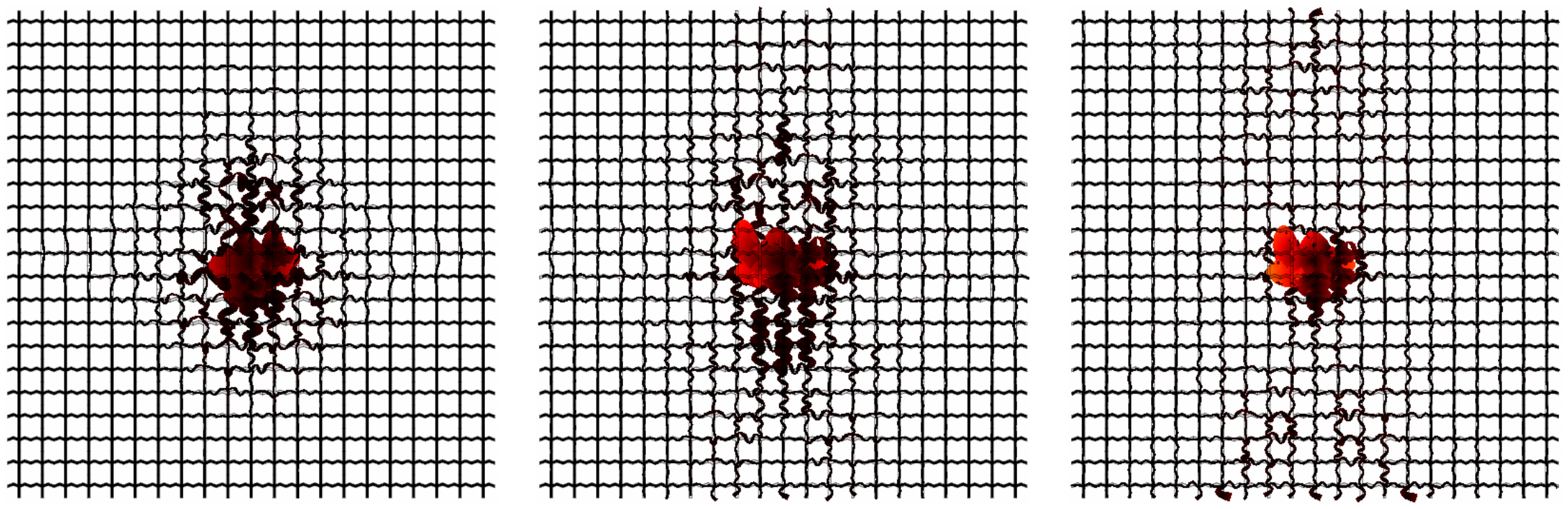

*t*=5T *t*=10T *t*=20T

(a) One *horizontal* (1, 0) point source

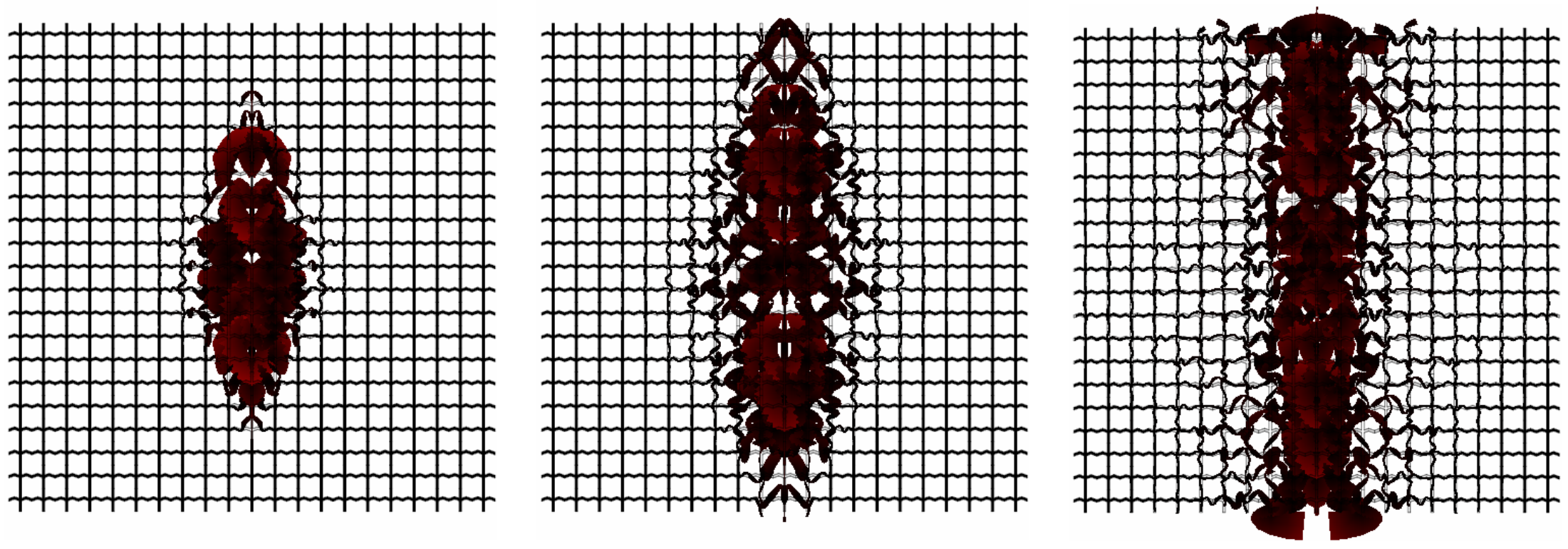

*t*=3T *t*=5T *t*=10T

(b) One *vertical* (0, 1) point source

Fig. 13 The transient response for one point source of (a) horizontal and (b) vertical polarization at $\Omega = 0.6$ for symmetric 1d-zigzag lattice with $T = 2\pi/\omega$.

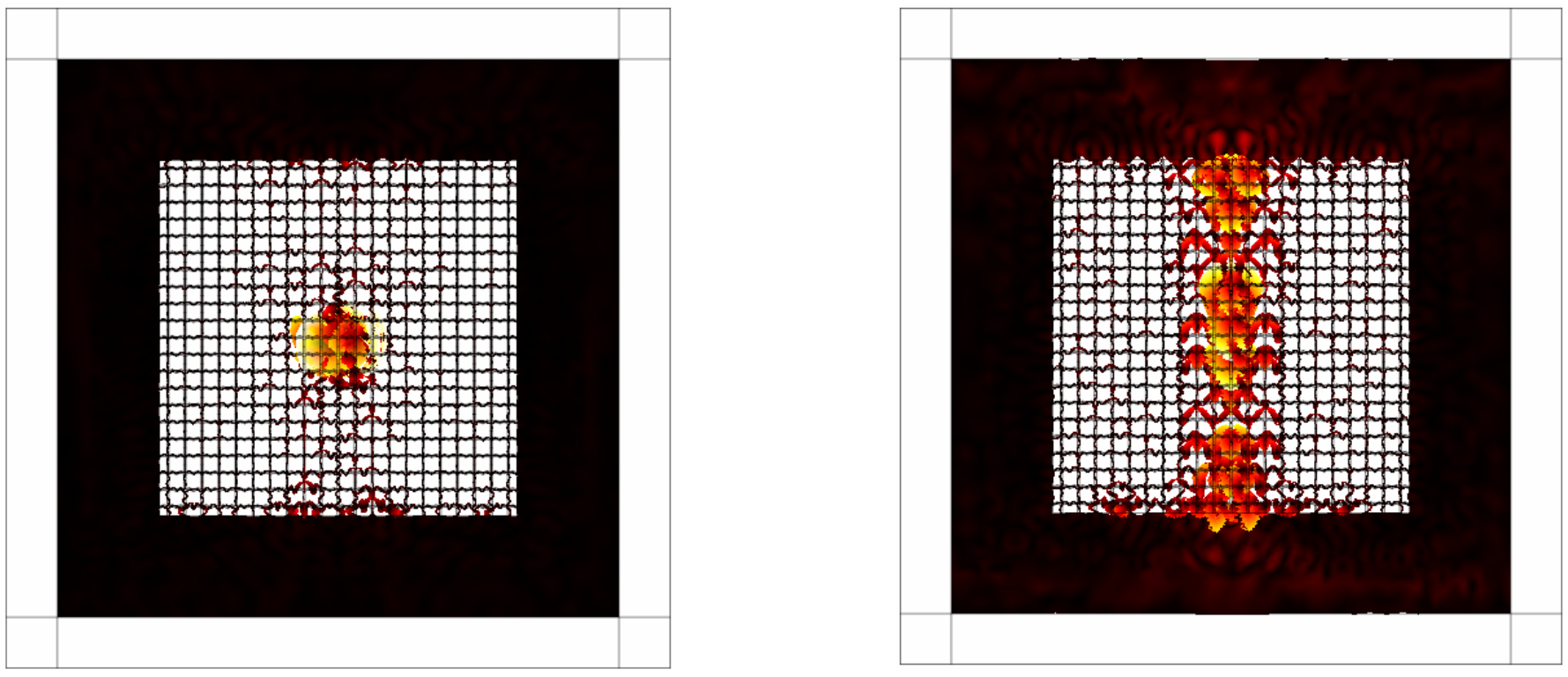

(a) One *horizontal* (1, 0) point source (b) One *vertical* (0, 1) point source

Fig. 14 The harmonic response for one point source of (a) horizontal and (b) vertical polarization at $\Omega = 0.6$ for symmetric 1d-zigzag lattice.

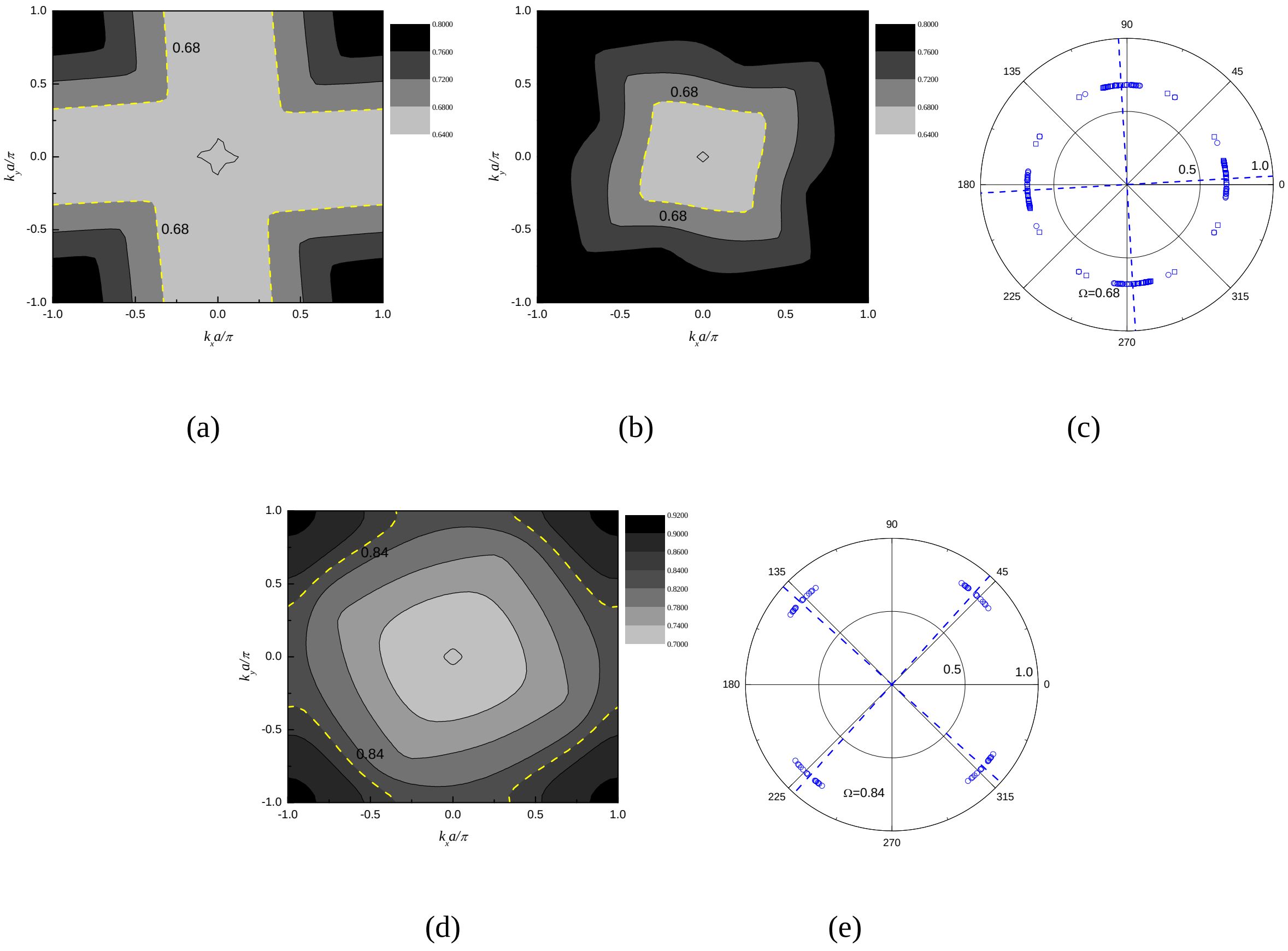

Fig. 15 Isofrequency curves of (a) the 7th and (b) 8th phase constant surface associated with the band structure in Fig. 8 for antisymmetric 2d-zigzag lattice. (c) Polar plot of the directional wave propagation at $\Omega = 0.68$, see the thick dashed line marked in Fig. 8. (d) Isofrequency curves of the 9th phase constant surface. (e) Polar plot of the directional wave propagation at $\Omega = 0.84$, see the thin dashed line marked in Fig. 8.

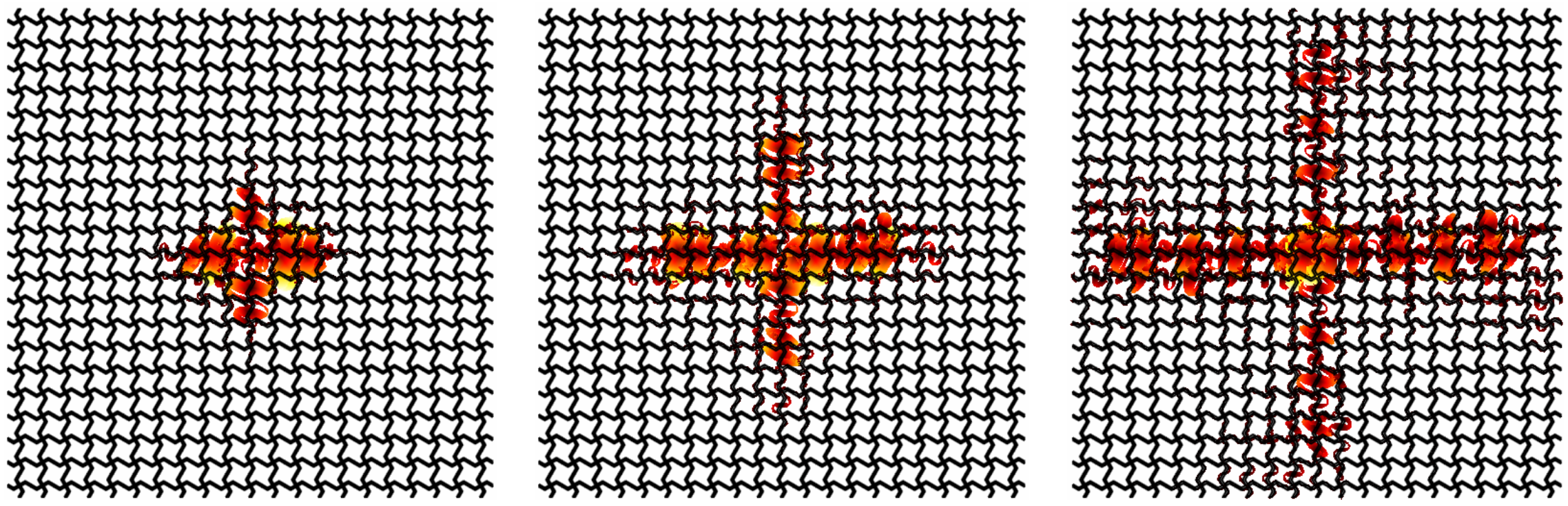

*t*=5T *t*=10T *t*=20T

(a) One *horizontal* (1, 0) point source

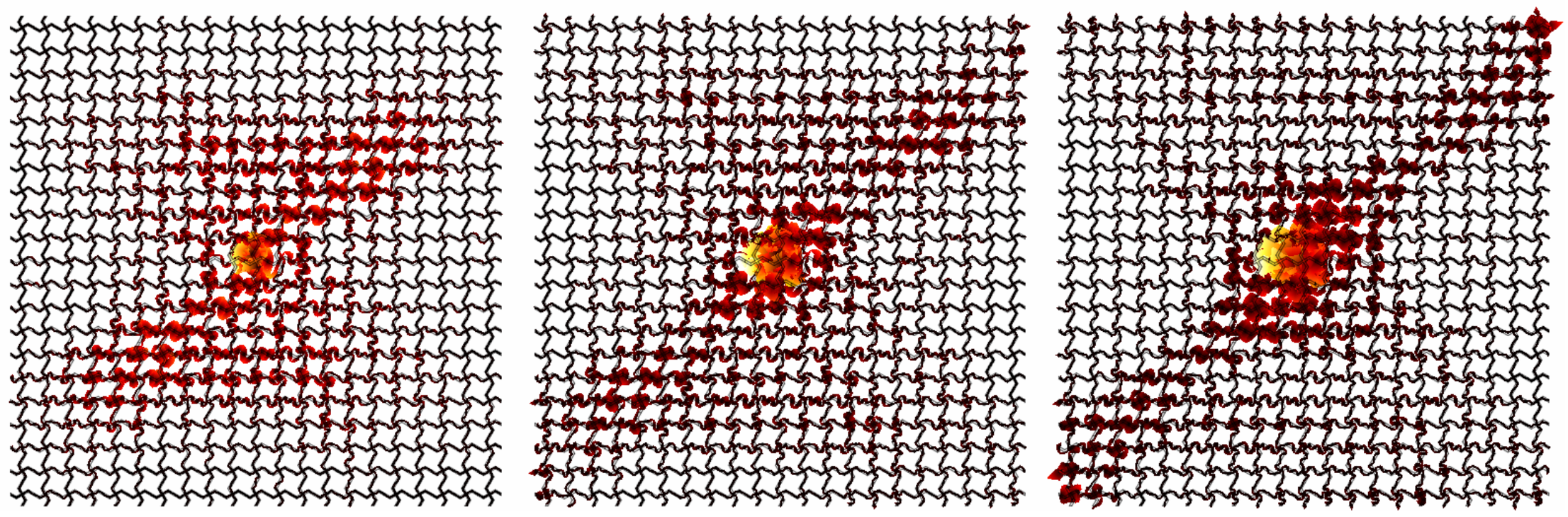

*t*=25T *t*=30T *t*=35T

(b) One *horizontal* (1, 0) point source

Fig. 16 The transient response at (a) $\Omega = 0.68$ and (b) $\Omega = 0.84$ for antisymmetric 2d-zigzag lattice.

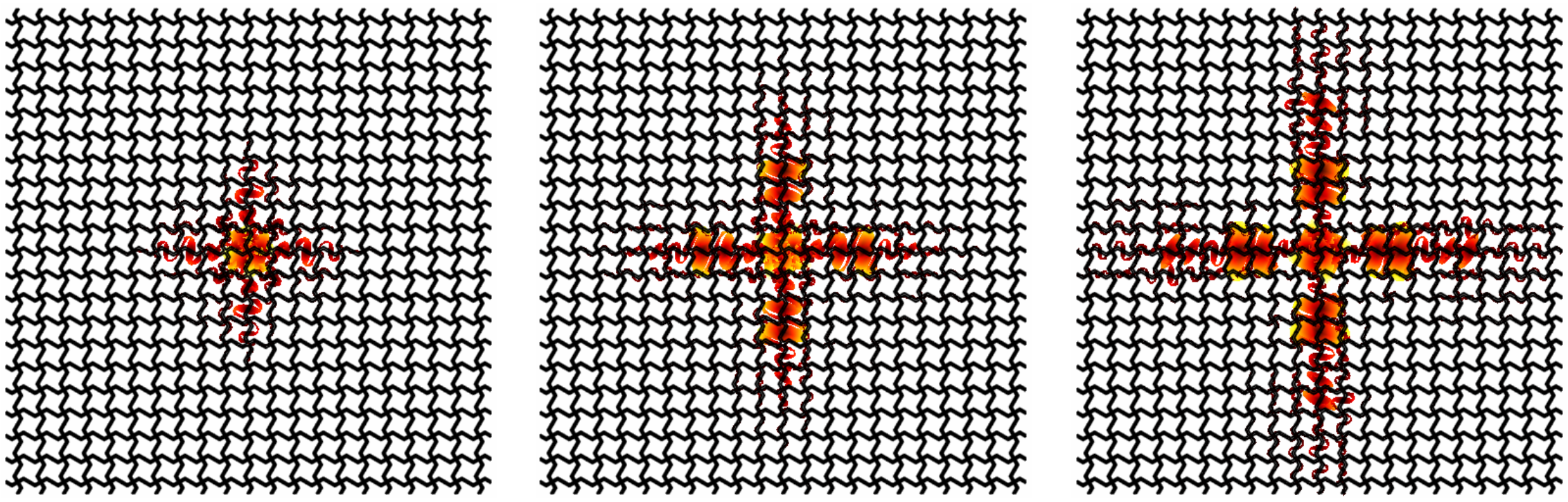

*t*=5T *t*=10T *t*=15T

Fig. 17 The transient response for a group of four point sources at $\Omega = 0.68$ for antisymmetric 2d-zigzag lattice.

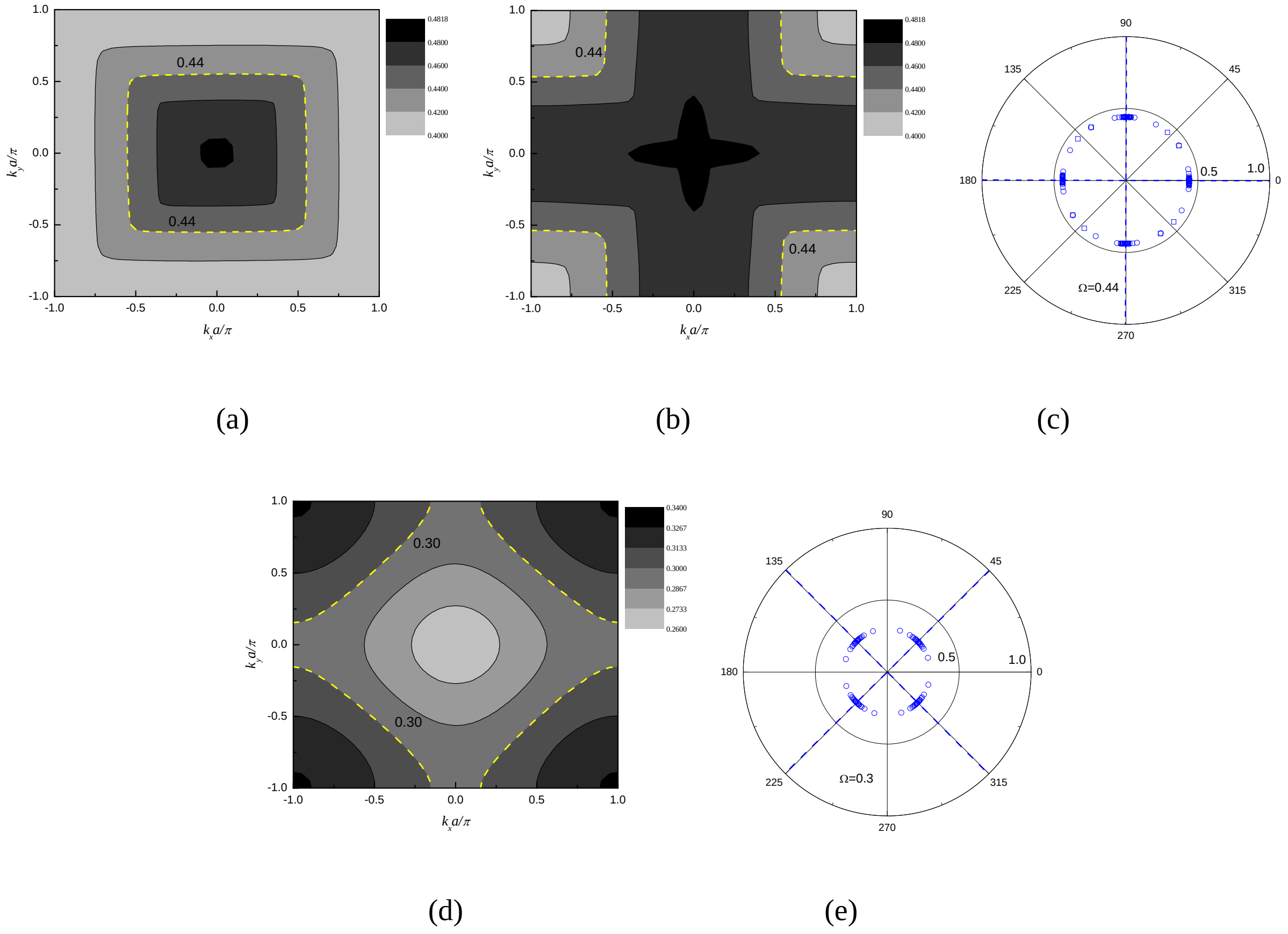

Fig. 18 Isofrequency curves of the (a) 6th and (b) 7th phase constant surface associated with the band structure in Fig. 10 for system with rotated cross holes. (c) Polar plot of the directional wave propagation at $\Omega = 0.44$, see the thick dashed line marked in Fig. 10. (d) Isofrequency curves of the 5th phase constant surface. (e) Polar plot of the directional wave propagation at $\Omega = 0.3$, see the thin dashed line marked in Fig. 10.

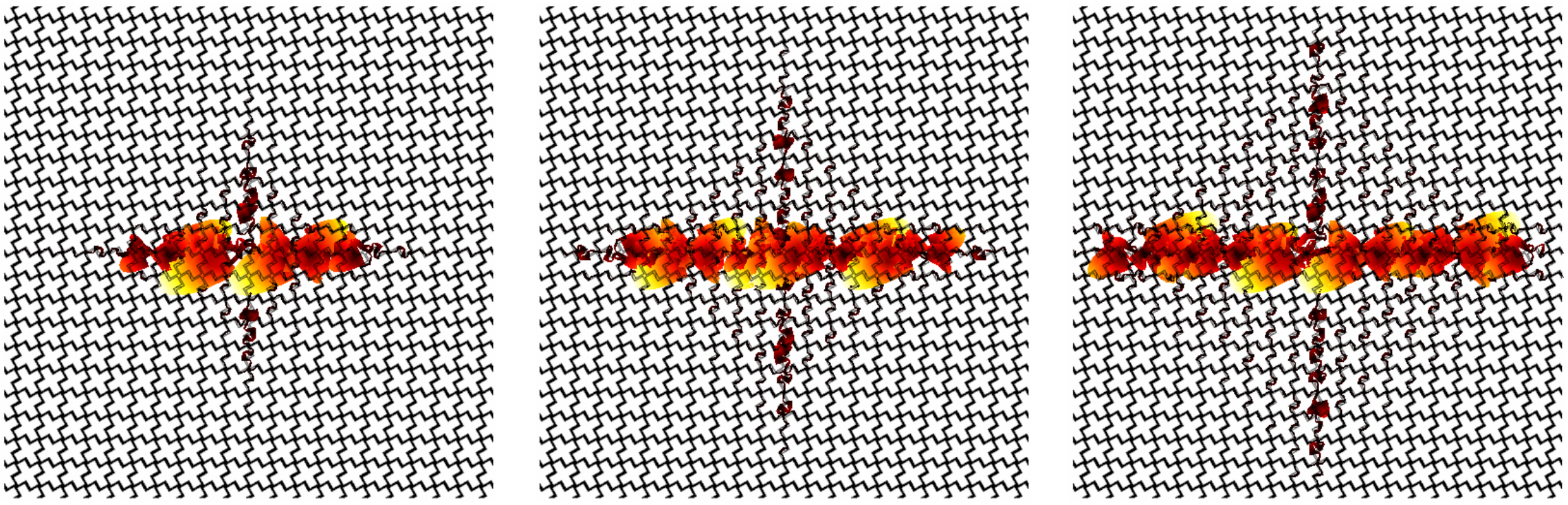

*t*=10T *t*=15T *t*=20T

(a) One *horizontal* (1, 0) point source

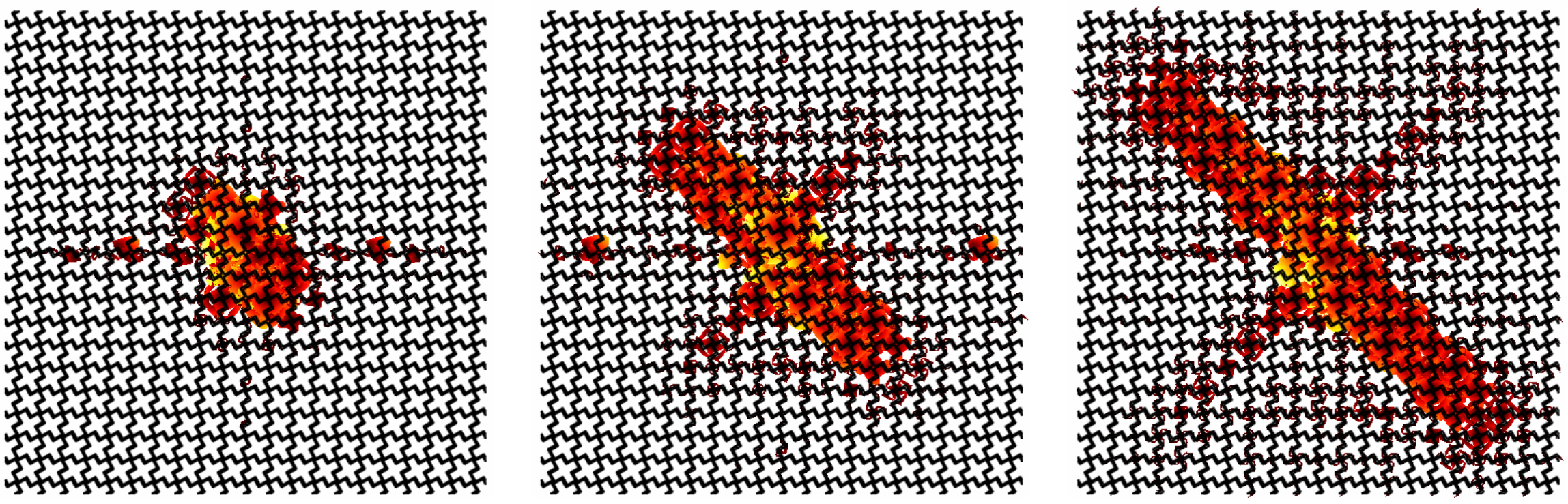

*t*=10T *t*=20T *t*=30T

(b) One *horizontal* (1, 0) point source

Fig. 19 The transient response at (a) $\Omega = 0.44$ and (b) $\Omega = 0.3$ for system with rotated cross holes.

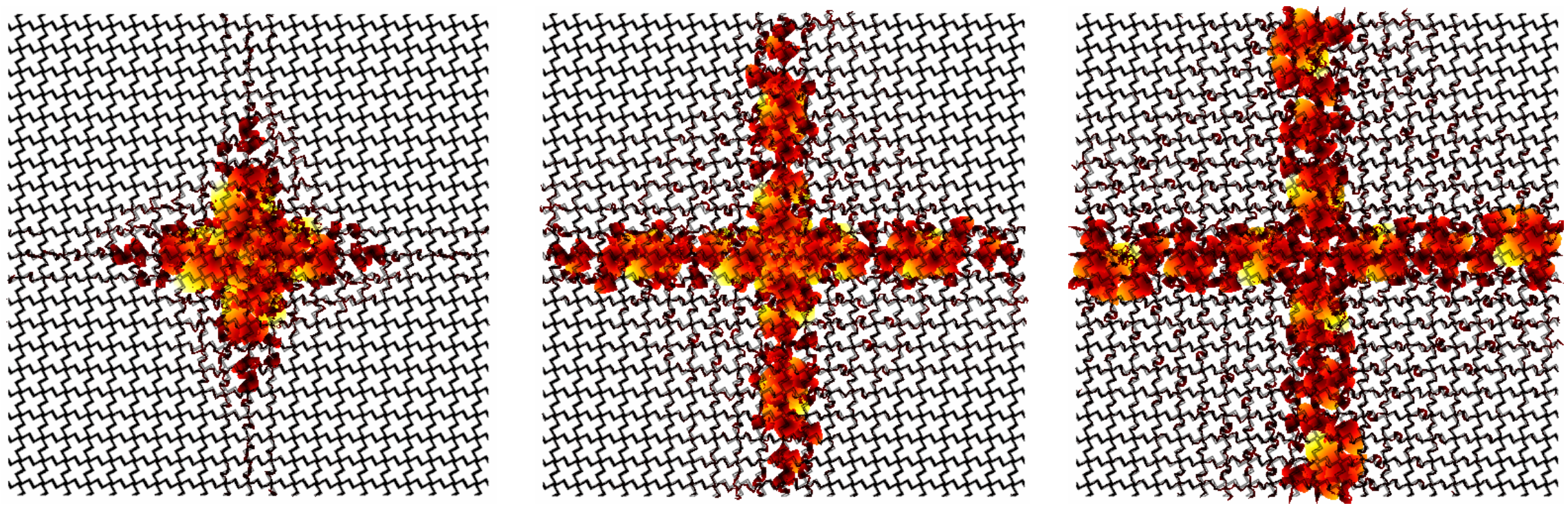

$t$=10T $t$=20T $t$=30T

Fig. 20 The transient response for a group of four point sources at $\Omega = 0.44$ for system with rotated cross holes.